\documentclass[11pt,a4paper,twoside]{article}
\pdfoutput=1
\usepackage{color}
\usepackage{hyperref}
\usepackage{ifpdf}
\ifpdf
\usepackage{graphicx}
\usepackage{epstopdf}
\else
\usepackage{graphicx}
\fi
\usepackage{float}
\usepackage{array}
\usepackage{placeins}
\usepackage{subfig}
\usepackage{amsmath}
\usepackage[sort&compress]{natbib}
\usepackage{rotating}
\usepackage{fullpage}
\usepackage{booktabs}
\usepackage{dcolumn}
\usepackage[english,british]{babel}
\usepackage[english]{varioref}
\usepackage{mciteplus}
\usepackage{setspace}
\usepackage{comment}
\usepackage{xspace}

\def \AT#1{\ensuremath{A_T^{\left(#1\right)}}\xspace}

\def \Afb{\ensuremath{A_{\mathrm{FB}}}\xspace}

\def \Fl{\ensuremath{F_\mathrm{L}}\xspace}

\def \S#1{\ensuremath{S_{#1}}\xspace}
\def \A#1{\ensuremath{A_{#1}}\xspace}

\def \kstar{{K^*}}

\def \eq{Eq.~}
\def \fig{Fig.~}
\def \figs{Figs.~}
\def \seq{Sec.~}
\def \tab{Tab.~}
\def \Rf{Ref.~}
\def \Refs{Refs.~}

\def \CL#1{\ensuremath{{\mathcal{C}_{#1}}}\xspace}
\def \CR#1{\ensuremath{{\CL{#1}^\prime}}\xspace}
\def \CLR#1{\ensuremath{\C{#1}^{(\prime)}}\xspace}

\def\lhcb {LHC{\em b\/}\xspace}
\def\btev {BTe\kern -0.1em V}

\def\babar{\mbox{\slshape B\kern-0.1em{\small A}\kern-0.1em B\kern-0.1em{\small
A\kern-0.2em R}}\xspace}
\def\lhc {LHC\xspace}
\def\belle {{\sc Belle}\xspace}

\def\mup        {\ensuremath{\mu^+}\xspace}
\def\mun        {\ensuremath{\mu^-}\xspace} 

\def\W      {\ensuremath{W}\xspace}

\def\s     {\ensuremath{s}\xspace}

\def\b     {\ensuremath{b}\xspace}

\def\kaon  {\ensuremath{K}\xspace}
\def\Kbar  {\kern 0.2em\overline{\kern -0.2em K}{}\xspace}

\def\Kz    {\ensuremath{K^0}\xspace}
\def\Kzb   {\ensuremath{\Kbar^0}\xspace}
\def\KzKzb {\ensuremath{\Kz \kern -0.16em \Kzb}\xspace}
\def\Kp    {\ensuremath{K^+}\xspace}
\def\Km    {\ensuremath{K^-}\xspace}

\def\KpKm  {\ensuremath{\Kp \kern -0.16em \Km}\xspace}

\def\Kstarz  {\ensuremath{K^{*0}}\xspace}
\def\Kstarzb {\ensuremath{\Kbar^{*0}}\xspace}
\def\Kstar   {\ensuremath{K^*}\xspace}
\def\Kstarb  {\ensuremath{\Kbar^*}\xspace}

\def\Dbar    {\kern 0.2em\overline{\kern -0.2em D}{}\xspace}

\def\Dz      {\ensuremath{D^0}\xspace}
\def\Dzb     {\ensuremath{\Dbar^0}\xspace}
\def\DzDzb   {\ensuremath{\Dz {\kern -0.16em \Dzb}}\xspace}
\def\Dp      {\ensuremath{D^+}\xspace}
\def\Dm      {\ensuremath{D^-}\xspace}

\def\DpDm    {\ensuremath{\Dp {\kern -0.16em \Dm}}\xspace}

\def\B       {\ensuremath{B}\xspace}
\def\Bbar    {\kern 0.18em\overline{\kern -0.18em B}{}\xspace}

\def\Bz      {\ensuremath{B^0}\xspace}
\def\Bzb     {\ensuremath{\Bbar^0}\xspace}
\def\BzBzb   {\ensuremath{\Bz {\kern -0.16em \Bzb}}\xspace}
\def\Bu      {\ensuremath{B^+}\xspace}
\def\Bub     {\ensuremath{B^-}\xspace}

\def\BpBm    {\ensuremath{\Bu {\kern -0.16em \Bub}}\xspace}
\def\Bd      {\ensuremath{B_d}\xspace}

\def\Bdb     {\ensuremath{\Bbar_d}\xspace}

\mathchardef\Upsilon="7107
\def\Y#1S{\ensuremath{\Upsilon{(#1S)}}\xspace}

\mathchardef\Deltares="7101
\mathchardef\Xi="7104
\mathchardef\Lambda="7103
\mathchardef\Sigma="7106
\mathchardef\Omega="710A

\def\Deltabar{\kern 0.25em\overline{\kern -0.25em \Deltares}{}\xspace}
\def\Lbar{\kern 0.2em\overline{\kern -0.2em\Lambda\kern 0.05em}\kern-0.05em{}\xspace}
\def\Sigbar{\kern 0.2em\overline{\kern -0.2em \Sigma}{}\xspace}
\def\Xibar{\kern 0.2em\overline{\kern -0.2em \Xi}{}\xspace}
\def\Obar{\kern 0.2em\overline{\kern -0.2em \Omega}{}\xspace}
\def\Nbar{\kern 0.2em\overline{\kern -0.2em N}{}\xspace}
\def\Xb{\kern 0.2em\overline{\kern -0.2em X}{}\xspace}

\def\BR         {{\ensuremath{\cal B}\xspace}}

\newcommand{\tev}{\ensuremath{\mathrm{\,Te\kern -0.1em V}}\xspace}
\newcommand{\gev}{\ensuremath{\mathrm{\,Ge\kern -0.1em V}}\xspace}
\newcommand{\mev}{\ensuremath{\mathrm{\,Me\kern -0.1em V}}\xspace}
\newcommand{\kev}{\ensuremath{\mathrm{\,ke\kern -0.1em V}}\xspace}
\newcommand{\ev}{\ensuremath{\mathrm{\,e\kern -0.1em V}}\xspace}
\newcommand{\gevc}{\ensuremath{{\mathrm{\,Ge\kern -0.1em V\!/}c}}\xspace}
\newcommand{\mevc}{\ensuremath{{\mathrm{\,Me\kern -0.1em V\!/}c}}\xspace}
\newcommand{\gevcc}{\ensuremath{{\mathrm{\,Ge\kern -0.1em V\!/}c^2}}\xspace}
\newcommand{\gevgevcccc}{\ensuremath{{\mathrm{\,Ge\kern -0.1em V^2\!/}c^4}}\xspace}
\newcommand{\mevcc}{\ensuremath{{\mathrm{\,Me\kern -0.1em V\!/}c^2}}\xspace}

\def\invfb   {\ensuremath{\mbox{\,fb}^{-1}}\xspace}

\def\mus  {\ensuremath{\rm \,\mus}\xspace}

\def\mus        {\ensuremath{\,\mu{\rm s}}\xspace}    

\def\to                 {\ensuremath{\rightarrow}\xspace}

\def\pep2{PEP-II}

\def\gsim{{~\raise.15em\hbox{$>$}\kern-.85em
          \lower.35em\hbox{$\sim$}~}\xspace}
\def\lsim{{~\raise.15em\hbox{$<$}\kern-.85em
          \lower.35em\hbox{$\sim$}~}\xspace}

\def\qsq                {\ensuremath{q^2}\xspace}
\def\thetaL                {\ensuremath{\theta_{l}}\xspace}
\def\thetaK                {\ensuremath{\theta_{\Kstar}}\xspace}

\def\CP                {\ensuremath{C\!P}\xspace}

\xspace

\newcommand{\app}       [1]  {{Acta Phys.\ Polon.\ {\bf #1}}}

\def\evtgen     {\mbox{\textsc{EvtGen}}\xspace}

\def\BdKsmm{\ensuremath{\Bd \to \Kstarz\mup\mun}\xspace}
\def\BdbKsmm{\ensuremath{\Bdb \to \Kstarzb\mup\mun}\xspace}

\def\BdbKsmmFull{\ensuremath{\Bdb \to \Kstarzb(\to\kaon\pi)\mup\mun}\xspace}

\def\Ceff#1{\ensuremath{\mathcal{C}_{#1}^{\mathrm{eff}}}}

\DeclareMathSizes{11}{11}{4}{9}

\newcolumntype{d}[1]{D{.}{.}{#1}}
\bibpunct{[}{]}{,}{n}{}{,}

\def \azeL{{A_0^L}}
\def \azeR{{A_0^R}}
\def \apaL{{A_\parallel^L}}
\def \apaR{{A_\parallel^R}}
\def \apeL{{A_\perp^L}}
\def \apeR{{A_\perp^R}}
\def \re{\text{Re}}
\def \im{\text{Im}}
\def \kstar{\ensuremath{K^{*}}\xspace}

\def \AT#1{\ensuremath{A_T^{\left(#1\right)}}\xspace}

\def \Afb{\ensuremath{A_{\mathrm{FB}}}\xspace}

\def \Fl{\ensuremath{F_\mathrm{L}}\xspace}

\def \eq{Eq.~}
\def \eqs{Eqs~}
\def \fig{Fig.~}
\def \figs{Figs~}
\def \seq{Sec.~}
\def \seqs{Secs~}
\def \app{App.~}
\def \tab{Tab.~}
\def \Rf{Ref.~}
\def \Refs{Refs~}

\def \CL#1{\ensuremath{C_{#1}}\xspace}
\def \CR#1{\ensuremath{\CL{#1}^\prime}\xspace}
\def \CLR#1{\ensuremath{\CL{#1}^{(\prime)}}\xspace}
\def \Ceff#1{\ensuremath{C^{\mathrm{eff}}_{#1}}\xspace}
\def \Ceffp#1{\ensuremath{C^{\prime\,\mathrm{eff}}_{#1}}\xspace}
\def \Ceffpn#1{\ensuremath{C^{(\prime)\,\mathrm{eff}}_{#1}}\xspace}

\def \rate#1{\ensuremath{\left<#1\right>_{\text{1-6}\gev^2}}\xspace}

\long\def\symbolfootnote[#1]#2{\begingroup\def\thefootnote{\fnsymbol{footnote}}\footnote[#1]{#2}\endgroup}

\begin{document}
\title{Constraining new physics with\\ 
$B\to K^*\mu^+\mu^-$ in the early \lhc era}
\author{Aoife Bharucha\thanks{a.k.m.bharucha@durham.ac.uk}\\
{\em University of Durham} \\ \\
William Reece\thanks{w.reece06@imperial.ac.uk}\\
{\em Imperial College London}
}

\begin{titlepage}
\begin{flushright}
\begin{tabular}{l}
IPPP/10/12\\
DCPT/10/24\\
IC/HEP/010-1
\end{tabular}
\end{flushright}
\vskip 1.5cm
\begin{center}
{\Large \bf \boldmath Constraining new physics with
$B\to K^*\mu^+\mu^-$ \\ [5pt] in the early \lhc era}
\vskip 1.3cm 
{\sc
Aoife Bharucha\symbolfootnote[1]{\href{mailto:a.k.m.bharucha@durham.ac.uk}{a.k.m.bharucha@durham.ac.uk}}$^{,1}$ and William Reece\symbolfootnote[2]{\href{mailto:will.reece@cern.ch}{will.reece@cern.ch}}$^{,2}$
} \vskip 0.5cm
$^1$ {\em IPPP, Department of Physics,
University of Durham, Durham DH1 3LE, UK}\\
\vskip 0.4cm
$^2$ {\em  Blackett Lab, Physics Department, Prince Consort Road, London SW7 2AZ, UK}\\

\vskip 1.5cm

{\large\bf Abstract\\[10pt]} \parbox[t]{\textwidth}{
\noindent We investigate the observables available in the angular distribution of $B\to K^*\mu^+\mu^-$ to identify those suitable for measurements in the first few years of \lhc data taking.
As experimental uncertainties will dominate, we focus on observables that are simple to measure, while maximizing the potential for discovery. There are three observables that may be extracted by counting signal events as a function of one or two decay angles and correspond to large features of the full angular distribution in the Standard Model: \Afb, \Fl, and \S{5}. Two of these are well known in the experimental community; however, we show that measuring \S{5} adds complementary sensitivity to physics beyond the Standard model. Like \Afb, it features a zero--crossing point with reduced hadronic uncertainties at leading order and in the large recoil limit. We explore the experimental sensitivity to this point at \lhcb and show that it may be measured with high precision due to the steepness of the \S{5} distribution. Current experimental model independent constraints on parameter space are presented and predictions made for the values of the \Afb and \S{5} zero--crossing points. The relative impact of \lhcb measurements of \Afb, \Fl, and \S{5}, with 2\invfb of integrated luminosity, is assessed. These issues are explored with a new model of the decay that can be used with standard simulation tools such as \evtgen.
}

\end{center}
\vskip 1.5cm
\begin{flushleft}
\begin{tabbing}
\= Keywords \=: \= \B-Physics; Beyond Standard Model; Rare Decays.\\ 
\end{tabbing}
\end{flushleft}

\end{titlepage}

\newcommand\T{\rule{0pt}{2.5 ex}}
\newcommand\Bo{\rule[-1.5ex]{0pt}{0pt}}
\newpage

\vspace{1in}

\section{Introduction}\label{sec:introduction}

The decay \BdbKsmm is a golden channel for the study of flavour changing neutral currents (FCNC) at the Large Hadron Collider (\lhc). The four-body final state, as $\Kstarzb \to \kaon \pi$, means that there is a wealth of information in the full-angular distribution that is complementary to that available in the widely studied $b \to s\gamma$ decays. In the presence of physics beyond the Standard Model (SM), new heavy degrees of freedom may enter the $\b \to \s$ loops. These can alter the decay amplitudes, affecting the full-angular distribution observed. This makes \BdbKsmm one of the most promising places in the flavour sector to search for new physics (NP) at the \lhc (see \Rf\cite{Hurth:2007xa} for a review). We concentrate on the large-recoil regime, where the energy of the \kstar is large such that QCD factorization is applicable. The low-recoil regime was described in \Rf\cite{Grinstein:2004vb}, however at present form factors in this regime are not well known. A number of interesting measurements have already been made \cite{Ishikawa:2003cp,Ishikawa:2006fh,Aubert:2006vb,Aubert:2008ju,Aubert:2008ps,Belle:2009zv,cdf_hcp2009_ksmm}. They are broadly in agreement with SM predictions; however, experimental precision is currently too low for firm conclusions to be drawn.

The properties of the full-angular distribution have been studied by many authors and a number of potential measurements have been identified; e.g. \Refs\cite{Ali:1991is,Burdman:1998mk, Kruger:2005ep,Egede:2008uy, Bobeth:2008ij, Altmannshofer:2008dz,Bharucha:2009ay}. Particular emphasis has been placed on finding angular observables with reduced theoretical uncertainties or enhanced sensitivity to particular classes of NP. However, in the first few years of \lhc data taking the dominant sources of uncertainty will be experimental; thus, the emphasis should be on finding quantities that can be cleanly measured with relatively small uncertainties. Once very large data sets have been collected, it will be possible to use a full-angular analysis to extract the various underlying amplitudes directly \cite{Egede:2008uy,Egede:2008}. This will allow the determination of many theoretically clean observables. However, performing this kind of analysis will not be possible until detectors are very well understood and the number of collected signal events are in the thousands. Prior to this, symmetries and asymmetries of the full-angular distribution can be used to extract some observables individually from angular projections \cite{Dickens:1045395,Jansen:1156131,Egede:1048970,Altmannshofer:2008dz,Bobeth:2008ij}.

In this paper, we focus on observables that correspond to large features in the  \BdbKsmm full-angular distribution and can be measured by counting the number of signal events as a function of one or two decay angles. We then investigate the relative experimental sensitivities to these observables at \lhcb \cite{Alves:2008zz} and their projected impact on the allowed parameter space after measurements with 2\invfb of integrated luminosity. The rest of the paper is structured as follows: In the next section we give a brief overview of the theoretical framework employed with details of the decay amplitude calculation; in \seq\ref{sec:NP}, observables that will be relevant for analyses with the first few years of \lhc data are discussed, and details of benchmark NP models provided. We also summarize the impact of existing experimental measurements on constraining the NP contribution to the Wilson coefficients. In \seq\ref{sec:sense}, we analyse the possibility of detecting NP effects at \lhcb using our chosen observables.  In \seq\ref{sec:impact}, the potential impact of these measurements on parameter space is assessed. Finally, in \seq\ref{sec:summary}, a short summary is given.

\vspace{2cm}

\section{Theoretical Details}\label{sec:theory}
\subsection{Introduction}\label{sec:theoryintro}

A decay model following \Rf\cite{Ali:1999mm} has become the standard tool for studies of \BdbKsmm within the experimental community due to its inclusion in the decay simulator \evtgen \cite{Lange:2001uf}. A significantly improved version of that model with much greater support for the simulation of NP as well as a state-of-the-art SM treatment has been developed as part of the present work \cite{evtgen-model}. We present our theoretical framework in a way that allows direct comparison with \Rf\cite{Ali:1999mm}, by expressing the decay amplitude in terms of the auxiliary functions used in that reference. Calculation of these requires Wilson coefficients, form factors and quantum-chromodynamics factorization (QCDF) corrections, as described in detail in this section.

\subsection{Wilson Coefficients}\label{sec:theory_wilson}
\begin{table}
\qquad\qquad
\begin{tabular}{|c|c|c|c|c|c|}
\hline
\T$\,\CL{1}(\mu)\quad$ &$\,\CL{2}(\mu)\quad$&$\,\CL{3}(\mu)\quad$\,&$\,\CL{4}(\mu)\quad$& $\,\CL{5}(\mu)\quad$&  $\,\CL{6}(\mu)\quad$\\
\hline
\,-0.135 \,&\, 1.054 \,&\, 0.012 \,&\, \,-0.033\, &\, 0.009 \,&\, -0.039 \,\\
\hline
\end{tabular}\\

\qquad\qquad\begin{tabular}{|c|c|c|c|}
\hline
\T $\,\Ceff{7}(\mu)$& $\,\Ceff{8}(\mu)$& $\,\Delta \Ceff{9}(\mu)$& $\,\Ceff{10}(\mu)\,$ \\
\hline
\, -0.306 \,&\, -0.159 \,&\,4.220 \,&\, -4.093 \,\\
\hline
\end{tabular}
\caption{\label{tab:wilson_mb_sm} SM Wilson coefficients at $\mu =
m_b = 4.52\gevcc$, where $\Delta \Ceff{9}(\mu) \sim \Ceff{9}(\mu)-Y(q^2) $.}
\end{table}
The Wilson coefficients, $\CL{i}(\mu)$, are process-independent coupling constants for the basis of effective vertices described by local operators, $\mathcal{O}_i(\mu)$, and encode contributions at scales above the renormalization scale, $\mu$. For a given NP model, new diagrams will become relevant and the $\CL{i}(\mu)$'s may change from their SM values; additional operators may also become important\footnote{A comprehensive review of effective field theories in weak decays can be found in \Rf\cite{Buchalla:1995vs}.}. The weak effective Hamiltonian, neglecting doubly Cabibbo-suppressed contributions, ${\cal H}_{\rm{eff}}^{(u)}$, is given by
\begin{equation} \label{eq:Heff}
    {\cal H}_{\mathrm{eff}}= - \frac{4\,G_F}{\sqrt{2}}
\lambda_t \left( \CL{1} \mathcal O^c_1 + \CL{2} \mathcal O^c_2 + \sum_{i=3}^{6} \CL{i} 
\mathcal O_i +\sum_{j} (\CL{j}\mathcal O_j + \CR{j} \mathcal
O'_j)\,\right),
\end{equation}
where $j=7,8,9,10,P,S$, $G_\mathrm{F}$ is the Fermi constant, and $\lambda_t=V_{tb}V_{ts}^*$ is the relevant combination of Cabibbo--Kobayashi--Maskawa (CKM) matrix elements. The operators $\mathcal O$ and $\mathcal O'$ are defined in \Rf\cite{Altmannshofer:2008dz}, and a subset is given explicitly in \app\ref{app:WCs}.

The primed operators have opposite chirality to the unprimed ones and their corresponding coefficients, $\CR{i}(\mu)$, are suppressed by $m_s/m_b$ or vanish in the SM; however, they may be enhanced by NP. We neglect the contributions from $\mathcal O_i^\prime$ for $1\leq i\leq 6$ as they are either heavily constrained by experimental results or generically small; NP contributions to $\mathcal O'_{7-10}$ may still be important and are included. We also include the scalar and pseudoscalar operators $\mathcal O^{(\prime)}_{S,P}$. These vanish in the SM but may arise in certain NP scenarios, for example in the case of an additional Higgs doublet.
 
 The Wilson coefficients are calculated by matching the full and effective
theories at the scale of the \W boson mass, $m_W$. For the SM Wilson coefficients, we aim at next-to-next-to-leading logarithmic (NNLL) accuracy. This requires calculating the matching conditions at $\mu=m_W$ to two-loop accuracy. This has been done in \Rf\cite{Bobeth:2004jz}. NP contributions are included to one-loop accuracy only, as two-loop corrections are expected to be small. This was shown explicitly for the MSSM in \Rf\cite{Bobeth:2001sq}. The Wilson coefficients must then be {\em evolved} down to the scale $\mu\sim m_b$. The evolution has been implemented using the full $10\times 10$ anomalous dimension matrix following \Refs\cite{Gambino:2003zm,Gorbahn:2004my,Gorbahn:2005sa}. The primed operators, $\mathcal O'_{7-10}$, are evolved as their unprimed equivalents; however, the scalar and pseudoscalar operators $\mathcal O^{(')}_{S/P}$ are defined to be conserved currents and do not mix with the other operators and so do not require evolution. For convenience, we define the following combinations of Wilson coefficients:
\begin{eqnarray}
\Ceff{7} & = & \frac{4\pi}{\alpha_s}\, \CL{7} -\frac{1}{3}\, \CL{3} -
\frac{4}{9}\,\CL{4} - \frac{20}{3}\, \CL{5}\, -\frac{80}{9}\,\CL{6};
\nonumber\\
\Ceff{8} & = & \frac{4\pi}{\alpha_s}\, \CL{8} + \CL{3} -
\frac{1}{6}\, \CL{4} + 20 \CL{5}\, -\frac{10}{3}\,\CL{6};
\nonumber\\
\Ceff{9} & = & \frac{4\pi}{\alpha_s}\,\CL{9} + Y(q^2);
\nonumber\\
\Ceff{10} & = & \frac{4\pi}{\alpha_s}\,\CL{10};\nonumber\\
\Ceffp{7,8,9,10} &=& \frac{4\pi}{\alpha_s}\,\CL{7,8,9,10}^{\prime}\,;\label{eq:WCseff}
\end{eqnarray} 
where \qsq is the invariant mass squared on the muon pair and $Y(\qsq)$ is defined in \Rf\cite{Beneke:2001at}. \tab\ref{tab:wilson_mb_sm} gives the values of the Wilson coefficients at $\mu = m_{b,{\rm PS}}(2{\rm GeV})$ in the SM. The treatment of quark masses in the PS scheme is discussed in \seq\ref{sec:numerics}.
 
\subsection{Form Factors}\label{sec:theory_form}

\BdbKsmm is characterized by eight form factors, $V(q^2)$, $A_{0-3}(q^2)$ and $T_{1-3}(q^2)$. These are hadronic quantities that, for certain ranges in $q^2$, may be obtained by non-perturbative methods. Their definition in terms of hadronic matrix elements can be found, for example, in \Rf\cite{Ball:2004rg}. Lattice field theory currently offers a prediction for the form factor $T_1(0)$ relevant to $B\to K^* \gamma$  \cite{Becirevic:2006nm}, but not for the others. However, QCD sum rules on the light cone (LCSR) is a well established alternative technique that provides results for the desired range in $q^2$ \cite{Ball:2004rg,Altmannshofer:2008dz}. It is an extension of classic QCD sum rules \cite{SVZ}, in which matrix elements are evaluated via both operator product expansion and dispersive representation. Quark-hadron duality then leads to sum rules for the desired hadronic quantities. LCSR follows a similar procedure to obtain sum rules for the form factors, but the operator product expansion in terms of vacuum condenstates is replaced by a light-cone expansion in terms of universal light-cone meson distribution amplitudes. A comprehensive review of QCD sum rules and LSCR can be found in \Rf\cite{Colangelo:2000dp}.

We use the full set of LCSR form factors in our model \cite{Ball:2004rg,BallFF:2007}, where the sum rules  for all form factors except for $A_0$ were calculated at $\mathcal{O}(\alpha_s)$ accuracy for twist-2 and-3 and tree-level accuracy for twist-4 contributions. Note that the normalization of the form factors we use differs slightly from \Rf\cite{Altmannshofer:2008dz}, however this will not have much impact on the observables, as they are normalized by the total decay rate, so the effect will cancel out. We estimate the uncertainties using the values provided in \Rf\cite{Ball:2004rg} for $q^2=0$, as shown in \tab\ref{tab:FFs}. Note that $A_3(0)$ and $T_2(0)$ are not included in the table, as they can be found using the relations $A_3(0)=A_0(0)$ and $T_2(0)=T_1(0)$.

  In the large energy limit of the \kstar, the form factors satisfy certain relations and, therefore, can be reduced to two {\em heavy-to-light} or {\em soft} form factors, denoted $\xi_\perp$ and $\xi_{\parallel}$ \cite{Charles:1998dr,Charles:1999gy,Dugan:1990de,Beneke:2000wa}. These reduced form factors are generally used within the QCDF framework \cite{Beneke:2001at,Beneke:2004dp}. The relations are studied through appropriate ratios of the LCSR predictions for the full form factors in Appendix B of \Rf\cite{Altmannshofer:2008dz}. It is shown that those involving $\xi_\perp$ are almost independent of $q^2$, but those involving  $\xi_{\parallel}$ have a definite dependence on $q^2$, so are probably more sensitive to the $1/m_b$ corrections neglected in QCDF.

\begin{table} \centering \begin{tabular}{|c|c|c|c|} \hline \T & $F(0)$
 & $\Delta_{\rm tot}$ & $\Delta_{a_1}$\\ \hline \hline \T $V$&  0.411
& 0.033 & 0.44 $\delta_{a_1}$\\ $A_0$ & 0.374 & 0.034 & 0.39
 $\delta_{a_1}$\\ $A_1$  &0.292 & 0.028 & 0.33 $\delta_{a_1}$\\ $A_2$
 & 0.259 & 0.027 & 0.31 $\delta_{a_1}$ \\ $T_1$  &0.333 & 0.028 & 0.34
 $\delta_{a_1}$\\ \Bo $T_3$ &0.202 & 0.018 & 0.18 $\delta_{a_1}$\\ 
 \hline \end{tabular} \caption{\label{tab:FFs} Form factors for
 \BdbKsmm from LCSR at $q^2=0$ \cite{Ball:2004rg}, as described in
 \seq\ref{sec:theory_form}. Here $\Delta_{\rm tot}$ is the total error
 arising from the uncertainty on all input parameters with the
 exception of the Gegenbauer moment $a_1$. $\Delta_{a_1}$ contains the
 uncertainty due to $a_1$, where $\delta_{a_1}$ is defined
 $\delta_{a_1}=a_1(\kstar,\, 1\mathrm{GeV})-0.1$.} \end{table}

\subsection{QCD Factorization Corrections}\label{sec:qcdf}
QCD factorization is a framework in which the $\mathcal{O}(\alpha_s)$ corrections to \BdbKsmm can be calculated in the combined heavy-quark and large-recoil energy limit; this  applies when the energy of the \kstar is large. These corrections take into account contributions that cannot be included in the form factors, such as the non-factorizable scattering effects arising from hard gluon exchange between the constituents of the \B meson.

Our calculation of the decay amplitude includes QCDF corrections at next-to-leading order (NLO) in $\alpha_s$ but leading order (LO)  in $1/m_b$. These corrections are included in the definitions of $\mathcal{T}_\parallel(q^2)$ and $\mathcal{T}_\perp(q^2)$ found in \Rf\cite{Beneke:2001at} and are given in terms of $\xi_\perp$ and $\xi_\parallel$; however, $\mathcal{O}(\alpha_s)$ factorizable corrections that arise from expressing the full form factors in terms of $\xi_\perp$ and $\xi_\parallel$ must then be subsumed. Following \Rf\cite{Altmannshofer:2008dz}, we instead express our LO results for the decay amplitude in terms of the full form factors. Factorizable corrections are then redundant and the main source of $\mathcal{O}(1/m_b)$ corrections is automatically included. In addition, we neglect weak annihilation corrections at LO in $1/m_b$ and $\mathcal{O}(\alpha_s)$ as they are dependent on the numerically small Wilson coefficients \CL{3} and \CL{4}.

We denote $\mathcal{T}^{\mathrm{NLO}}_\parallel(q^2)$ and $\mathcal{T}^{\mathrm{NLO}}_\perp(q^2)$ to be the analogues of $\mathcal{T}_\parallel(q^2)$ and $\mathcal{T}_\perp(q^2)$ from \Rf\cite{Beneke:2001at}  with the only relevant $\mathcal{O}(\alpha_s)$ contributions included. We also define $\mathcal{T}'^{\mathrm{NLO}}_\parallel(q^2)$ and $\mathcal{T}'^{\mathrm{NLO}}_\perp(q^2)$; the primes indicate that the unprimed Wilson coefficients should be replaced by their primed equivalents. In order to extend the results of \Rf\cite{Ali:1999mm} to include NLO corrections, we must make the following replacements:
\begin{align}
 \nonumber \Ceffpn{7} T_1(q^2) \quad \to& \quad \Ceffpn{7} T_1(q^2)+\mathcal{T}^{(\prime)\,\mathrm{NLO}}_\perp(q^2);\\
\nonumber \Ceffpn{7} T_2(q^2) \quad \to& \quad \Ceffpn{7} T_2(q^2)+2 \frac{E_{\kstar}(q^2)}{m_B} \mathcal{T}^{(\prime)\,\mathrm{NLO}}_\perp(q^2);\\
 \Ceffpn{7} T_3(q^2) \quad \to& \quad \Ceffpn{7} T_3(q^2)+\mathcal{T}^{(\prime)\,\mathrm{NLO}}_{\perp}(q^2)+\mathcal{T}^{(\prime)\,\mathrm{NLO}}_{\parallel}(q^2);
\end{align}
where $E_{\kstar}(q^2)$ is the energy of the \kstar and $m_B$ is the mass of the B meson.

We have now introduced the Wilson coefficients, form factors and defined the QCD factorization corrections. These are all ingredients for the auxiliary functions describing the decay amplitude, as seen in the following subsection.

\subsection{Decay Amplitude}\label{sec:decayamp}
The Hamiltonian defined in \eq(\ref{eq:Heff}), combined with the standard definitions of the form factors, leads to the following decay amplitude \cite{Ali:1999mm,Yan:2000dc}:
\begin{align}
\mathcal{M} &\propto \left[\mathcal{T}_\mu^1\left(\bar{\mu}\,\gamma^\mu\, \mu\right) +
\mathcal{T}_\mu^2\left(\bar{\mu}\,\gamma^\mu\gamma_5\, \mu\right) +
\mathcal{S}(\bar{\mu}\,\mu)\right]\label{eqn:decay_amp}\displaybreak[0]
\intertext{where}
\mathcal{T}_\mu^1 &= A(q^2)\epsilon_{\mu\rho\alpha\beta}\epsilon^{*\rho}\,\hat{p}^\alpha_\B\,\hat{p}^\beta_{\kstar} - 
iB(q^2)\,\epsilon^{*}_\mu + 
iC(q^2)(\epsilon^*\cdot\hat{p}_\B)\,\hat{p}_{\mu} + 
iD(q^2)(\epsilon^*\cdot\hat{p}_\B)\,\hat{q}_\mu\displaybreak[0]\\
\mathcal{T}_\mu^2 &= E(q^2)\epsilon_{\mu\rho\alpha\beta}\,\epsilon^{*\rho}\hat{p}^\alpha_\B\,\hat{p}^\beta_{\kstar} - 
iF(q^2)\,\epsilon^{*}_\mu + 
iG(q^2)(\epsilon^*\cdot\hat{p}_\B)\,\hat{p}_{\mu} + 
iH(q^2)(\epsilon^*\cdot\hat{p}_\B)\,\hat{q}_\mu\displaybreak[0]\label{eqn:decay_amp_T1T2S}
\intertext{and}
\mathcal{S} & = i2\hat{m}_{\kstar}(\epsilon^*\cdot\hat{p}_\B)\,I(q^2).
\end{align}
Here, $p_{\B,\kstar}$ and $m_{\B,\kstar}$ are the four-momenta and masses of the respective particles in the \B meson rest frame, \mbox{$p \equiv p_B + p_{\kstar}$}, \mbox{$q \equiv p_B-p_{\kstar}$}, and $\epsilon^*_\mu$ is the \kstar polarization vector. The circumflex denotes division by $m_B$ (e.g. $\hat{m}_{\kstar} \equiv m_{\kstar}/m_B$). The auxiliary functions $A$-$I(q^2)$ follow \Rf\cite{Ali:1999mm}; however, we have updated the previous expressions to include additional primed, scalar, and pseudoscalar operators, as well as QCDF correction via  $\mathcal{T}^{(')\rm NLO}_\parallel(q^2)$ and $\mathcal{T}^{(')\rm NLO}_\perp(q^2)$ as outlined in \seq\ref{sec:qcdf}. They are defined as:
\begin{subequations}
\label{eq:AmpAtoI}
\begin{align}
\nonumber A(q^2)=&\frac{2}{1+\hat{m}_{\kstar}} (\Ceff{9}+\Ceffp{9}) V(q^2)+\frac{4 \hat{m}_b }{\hat{q}^2} \bigg((\Ceff{7}+\Ceffp{7}) T_1(q^2)\\
&+{\mathcal T}^{\rm NLO}_{\perp}(q^2)+{\mathcal T}^{\prime\,\rm NLO}_{\perp}(q^2)\bigg);\displaybreak[0]\\
\nonumber B(q^2)=&(1+\hat{m}_{\kstar})\bigg\{(\Ceff{9}-\Ceffp{9}) A_1(q^2)+\frac{2\hat{m}_b}{\hat{q}^2}(1-\hat{m}_{\kstar})\bigg((\Ceff{7}-\Ceffp{7}) T_2(q^2)\\
  &+ 2 \hat{E}_{\kstar}(q^2)({\mathcal T}^{\rm NLO}_\perp(q^2)-{\mathcal T}^{\prime\,\rm NLO}_\perp(q^2))\bigg)\bigg\};\displaybreak[0]\\ 
\nonumber C(q^2)=&\frac{1}{1-\hat{m}_{\kstar}^2}\bigg\{(1-\hat{m}_{\kstar}) (\Ceff{9}-\Ceffp{9}) A_2(q^2)\\
\nonumber &+2 \hat{m}_b \bigg((\Ceff{7}-\Ceffp{7}) (T_3(q^2)+\frac{1-\hat{m}_{\kstar}^2}{\hat{q}^2} T_2(q^2))\\
\nonumber &+(1+\frac{(1-\hat{m}_{\kstar}^2)\,2 \hat{E}_{\kstar}(q^2)}{\hat{q}^2 })({\mathcal T}^{\rm NLO}_\perp(q^2)-{\mathcal T}^{\prime\,\rm NLO}_\perp(q^2)) \\
& +{\mathcal T}^{\rm{NLO}}_\parallel(q^2)-{\mathcal T}^{\prime\,\rm NLO}_\parallel(q^2)\bigg)\bigg\};\displaybreak[0]\\
E(q^2)=&\frac{2}{(1+\hat{m}_{\kstar})} (\Ceff{10}+\Ceffp{10}) V(q^2);\displaybreak[0]\\
F(q^2)=&(1+\hat{m}_{\kstar})(\Ceff{10}-\Ceffp{10}) A_1(q^2);\displaybreak[0]\\
G(q^2)=&(\Ceff{10}-\Ceffp{10}) \frac{A_2(q^2)}{(1+\hat{m}_{\kstar})};\displaybreak[0]\\
\nonumber H(q^2)=&\frac{1}{\hat{q}^2} (\Ceff{10}-\Ceffp{10}) \bigg((1+\hat{m}_{\kstar})A_1(q^2)-(1-\hat{m}_{\kstar})A_2(q^2)\\
& -2 \hat{m}_{\kstar} A_0(q^2)\bigg)-\frac{\hat{m}_{\kstar} m_B}{2 \hat{m}_\mu} A_0(q^2)  (\CL{P}-\CR{P});\displaybreak[0]\\
I(q^2)=&- A_0(q^2) (\CL{S}-\CR{S}).
\end{align}
\end{subequations}
The recoil energy of the \kstar is given by
\begin{equation}
 E_{\kstar}(q^2)=\frac{m_B^2+m_{\kstar}^2-q^2}{2 m_B}.
\end{equation}
Using the equations of motion for the muons,
\begin{equation}
q^\mu(\bar{\mu}\gamma_\mu\, \mu) = 0 \qquad {\rm and} \qquad q^\mu(\bar{\mu}\gamma_\mu\gamma_5\, \mu)= -2m_\mu \bar{\mu}\gamma_5\, \mu,
\end{equation}
where $m_\mu$ is the muon mass, we see that $D(q^2)$ vanishes and $H(q^2)$ is
suppressed by a power of $m_\mu$. However, $H(q^2)$ receives a pseudoscalar contribution inversely proportional to $m_\mu$ allowing for some sensitivity to $\CL{P}-\CR{P}$ \cite{Yan:2000dc}. 
 The observables described in \seq\ref{sec:observables} (e.g. \eqs(\ref{eqn:obsS})--(\ref{eqn:obsA})) may be calculated directly from the amplitudes given in \eq(\ref{eq:AmpAtoI}); the necessary formulae are presented in \app\ref{app:Is} and implemented in our model.
 
 \begin{table}
\centering
\begin{tabular}{|c|c|c|}
\hline
\T\,Parameter \,& Value &\, \Rf\,\\
\hline
\hline
\T$m_s$&0.104&\cite{Amsler:2008zz}\\
\,$m_{c,{\rm PS}}(0.7{\rm\, GeV})$\, & 1.5 GeV &\cite{Signer:2008da}\\
$m_{b,{\rm PS}}(2{\rm\,GeV})$ & 4.52 GeV &\cite{Pineda:2006gx} \\
\Bo$\hat{m}_t(\hat{m}_t)$ &\, 162.3 GeV\, & \cite{CDF:2008vn} \\
\hline
\end{tabular}
\caption{\label{tab:QuarkMasses} Quark masses}
\end{table} 
\subsection{Numerical Input}\label{sec:numerics}
\subsubsection{Quark Masses}\label{sec:quarkmasses} The calculation of the auxiliary functions requires the bottom quark pole mass, which is known to contain large long-distance corrections. To avoid this, a renormalization scheme, known as the potential subtraction scheme (PS), was introduced in \Rf\cite{Beneke:1998rk}. The quark mass defined in the PS scheme has the advantage that the large infrared contributions are absent, while being numerically close to the pole mass. It is suitable for calculations in which the quark is nearly on-shell. Following \Rf\cite{Beneke:2001at}, we replace the pole mass by the PS mass, $m_\mathrm{PS}(\mu_f)$, using 
\begin{equation}\label{eq:PStoPole}
m=m_\mathrm{PS}(\mu_f)+\frac{4 \alpha_s}{3 \pi} \mu_f+\mathcal{O}(\alpha_s^2)
\end{equation}
and neglect any resulting terms of $\mathcal{O}(\alpha_s^2)$. Here $\mu_f$ is the scale at which the PS mass is calculated. All
occurrences of the symbol $m_{b}$ in our formulae refer to the PS mass, $m_{b,{\rm PS}}(2{\rm\,GeV})$, as shown in \tab\ref{tab:QuarkMasses}.

The operator $\mathcal{O}_7$ is defined in terms of the modified minimal subtraction ($\mathrm{\overline{MS}}$) mass. In the $\mathrm{\overline{MS}}$ scheme, the $1/\epsilon$ poles are simply removed, along with the associated terms in $\gamma$ and $4\pi$. Therefore, when the $b$ quark mass arises in combination with $\Ceff{7}$, we replace the $\mathrm{\overline{MS}}$ mass, $\bar{m}$, by the pole mass, using 
\begin{equation}
\bar{m}(\mu)=m\left(1+\frac{\alpha_s}{3 \pi}\left(3 \ln\frac{m_b^2}{\mu^2}-4\right)+\mathcal{O}(\alpha_s^2)\right).
\end{equation}
This leads to factorizable $\mathcal{O}(\alpha_s)$ corrections to  $\mathcal{T}^{\rm{NLO}}_{\perp/\parallel}(q^2)$ and $\mathcal{T}'^{\rm{NLO}}_{\perp/\parallel}(q^2)$ as found in \Rf\cite{Beneke:2001at}. 

For consistency, we calculate the charm quark pole mass using \eq(\ref{eq:PStoPole}).  Here the PS mass is taken from the most recent calculation as in \tab\ref{tab:QuarkMasses}. The resulting pole mass agrees with results in \Rf\cite{Amsler:2008zz}, where it is calculated from the $\overline{\rm{MS}}$ mass. The top quark mass enters the calculation of the Wilson coefficients, and for this we use the $\overline{\rm{MS}}$ mass in \tab\ref{tab:QuarkMasses}, as in \Rf\cite{Altmannshofer:2008dz}.
\begin{table}
\centering
\begin{tabular}{|c|c|c||c|c|c|}
\hline
\T\,Parameter\, & Value &\,\Rf\,& Parameter & Value &\,\Rf\, \\
\hline
\hline
\T$f_B$ & $200\pm 25$ MeV &\cite{Onogi:2006km} &$a_{1,K^*}^\perp$(2 GeV) & $0.03\pm 0.03$& \cite{Ball:2007zt}\\
$\lambda_B$(2.2 GeV) & \,$0.51 \pm 0.12$ GeV \,& \cite{Ball:2006nr} & $a_{1,K^*}^{\parallel}$(2 GeV) & $0.02\pm 0.02$ & \cite{Ball:2007zt}\\
$f_{K^*}^\perp$(2 GeV) & $163\pm 8 $ MeV &\cite{Ball:2007zt} &
 $a_{2,K^*}^\perp$(2 GeV) & $0.08\pm 0.06$&\cite{Ball:2007zt}\\
\Bo $\,f_{K^*}^{\parallel}\,$ & $220 \pm 5$ MeV &\cite{Ball:2007zt} & $\,a_{2,K^*}^\parallel\,$(2 GeV) & $0.08\pm 0.06$ &\cite{Ball:2007zt}\\  

\hline
\end{tabular}
\caption{\label{tab:Hadronic}Hadronic parameters}
\end{table} 
\subsubsection{Hadronic Parameters}\label{sec:hadronicparams} In addition to the form factors described in \seq\ref{sec:theory_form}, the QCDF corrections require light-cone distribution amplitudes and decay constants. The light-cone distribution amplitude for both the \B and \kstar mesons enter the hard scattering corrections. For the \B meson we follow the prescription in \Rf\cite{Beneke:2001at} using the values for $\Lambda_B$ given in \tab\ref{tab:Hadronic}.
 For the \kstar meson we use the standard Gegenabauer expansion,
\begin{equation}
\Phi_{K^*}^m=6 u(1-u)(1+a_{1,K^*}^m C_1^{(3/2)}(2u-1)+a_{2,K^*}^m C_2^{(3/2)}(2u-1)),
\end{equation} 
for $m=\perp,\parallel$, taking the coefficients from \tab\ref{tab:Hadronic}.
We also require the decay constants for both the \B and \kstar mesons. Additional parameters are summarized in \tab{\ref{tab:BasicConstants}}.

\section{Observables and New physics}\label{sec:NP}
\begin{table}
\centering
\begin{tabular}{|c|c||c|c|}
\hline
\T\,Parameter\, & Value  & \,Parameter\, & Value \\
\hline
\hline
\T$m_B$ & 5.28 GeV&$V_\mathrm{us}$&$0.226\pm 0.002$\\
$m_{K^*}$ &\, 0.896 GeV\,&$V_\mathrm{ub}$ &\, $(3.93\pm 0.36) 10^{-3}$\,\\
$m_\mu$ & 0.106 GeV & $\gamma$& $(77^{+30}_{-32})^\circ $\\
\Bo$M_W$ & 80.4 GeV & $G_F$ & $(1.166)10^{-5}\,\mathrm{GeV}^{-2}$ \\
\hline
\end{tabular}
\caption{\label{tab:BasicConstants} CKM matrix parameters, additional masses and constants from \Rf\cite{Amsler:2008zz}.}
\end{table}

Having established the basic theoretical framework, we proceed to discuss experimental observables for \BdbKsmm.

\subsection{Observables}\label{sec:observables}
The full-angular decay distribution can be written as:
\begin{equation}\label{eq:d4Gamma}
  \frac{\rm{d}^4\Gamma}{\rm{d}q^2\, \rm{d}\cos\theta_l\,\rm{d}\cos\theta_{K^*}\, \rm{d}\phi} =
   \frac{9}{32\pi} I(q^2, \theta_l, \theta_{K^*}, \phi),
\end{equation}
where the angles \thetaK, \thetaL and $\phi$ are defined as follows: \thetaK is the angle between the $K^-$ and $\bar{B}$ in the rest frame of the \Kstarb, and is defined in the range $-1\leq \cos\thetaK\leq 1$; \thetaL is defined as the angle between the $\mu^-$ and $\bar{B}$ in the di-muon centre of mass frame, and is defined in the range $-1\leq \cos\theta_l\leq 1$; $\phi$ is the angle between the normal to the $K$-$\pi$ plane and the normal to the di-muon plane, and is defined in the range $0\leq\phi\leq 2 \pi$. For the conjugate decay, the angles are defined analogously, but with reference to the $K^+$ and $\mu^+$. We can then express $I(q^2, \theta_l, \theta_{K^*}, \phi)$ in terms of these angles as follows:
\begin{align} \label{eq:angulardist}
  I(q^2, \theta_l, \theta_{K^*}, \phi)& = 
      I_1^s \sin^2\theta_{K^*} + I_1^c \cos^2\theta_{K^*}
      + (I_2^s \sin^2\theta_{K^*} + I_2^c \cos^2\theta_{K^*}) \cos 2\theta_l
\nonumber \\       
    & + I_3 \sin^2\theta_{K^*} \sin^2\theta_l \cos 2\phi 
      + I_4 \sin 2\theta_{K^*} \sin 2\theta_l \cos\phi 
\nonumber \\       
    & + I_5 \sin 2\theta_{K^*} \sin\theta_l \cos\phi
\nonumber \\      
    & + (I_6^s \sin^2\theta_{K^*} +
      {I_6^c \cos^2\theta_{K^*}})  \cos\theta_l 
      + I_7 \sin 2\theta_{K^*} \sin\theta_l \sin\phi
\nonumber \\ 
    & + I_8 \sin 2\theta_{K^*} \sin 2\theta_l \sin\phi
      + I_9 \sin^2\theta_{K^*} \sin^2\theta_l \sin 2\phi .
\end{align}
The angular coefficients $I_i^{(a)}$, where $i=1$ to $9$ and $a=s$ or $c$, describe the decay distribution. A natural set of observables was identified in \Rf\cite{Altmannshofer:2008dz} by taking combinations of these $I_i^{(a)}$'s that emphasize \CP-conserving and \CP-violating effects. These were defined as 
\begin{align}
S^{(s/c)}_i &=(I^{(s/c)}_i+\bar{I}^{(s/c)}_i)\Bigg/\frac{\rm{d}(\Gamma+\bar{\Gamma})}{\rm{d}q^2} \label{eqn:obsS},\\
A^{(s/c)}_i &=(I^{(s/c)}_i-\bar{I}^{(s/c)}_i)\Bigg/\frac{\rm{d}(\Gamma+\bar{\Gamma})}{\rm{d}q^2} \label{eqn:obsA},
\end{align} 
where the $A^{(s/c)}_i$'s have also been studied in \Rf\cite{Bobeth:2008ij}. We introduce the rate average, which, for a variable $V(\qsq)$, is given by
\begin{equation}
\rate{V} = \int^{6\gev^2}_{1\gev^2}\mathrm{d}\qsq\left( V(\qsq)\,\frac{\rm{d}(\Gamma+\bar{\Gamma})}{\rm{d}q^2}\right) \Bigg/ \int^{6\gev^2}_{1\gev^2}\mathrm{d}\qsq\frac{\rm{d}(\Gamma+\bar{\Gamma})}{\rm{d}q^2}.\label{eqn:vint}
\end{equation}
Using \eq(\ref{eqn:obsS}), it is possible to reconstruct standard observables such as the forward-backward asymmetry, \Afb,  and the longtitudinal polarization fraction, \Fl:
\begin{equation}
 \Afb=\frac{3}{8}(2 S_6^s+S_6^c)\qquad \qquad \rm{and} \qquad \qquad \Fl= -S_2^c.
\end{equation}

As explained in \seq{\ref{sec:introduction}, our focus is on those observables that will be measurable at \lhcb without a full-angular analysis. In order to keep the experimental complexity to a minimum, these observables should require information on only one or two of the angles. \Afb, which depends only on $\theta_l$, and \Fl, which depends only on $\theta_{\kstar}$, are well known examples. They can be expressed as:
\begin{align}
\Afb=&\frac{4}{3}\left( \int_{0}^{1}-\int_{-1}^{0}\right)\rm{d}\theta_l\frac{\mathrm{d}^2(\Gamma+\bar{\Gamma})}{ d\mathrm{q}^2\,\rm{d}\theta_l}\bigg/\frac{\mathrm{d}(\Gamma+\bar{\Gamma})}{\mathrm{d}q^2}; \label{eqn:afb_counting}\displaybreak[0]\\
\Fl=&\frac{1}{9}\left( 16 \int_{-1/2}^{1/2}\frac{\mathrm{d}(\Gamma+\bar{\Gamma})}{ d\mathrm{q}^2\,\rm{d}\cos\theta_{\kstar}}\bigg/\frac{\mathrm{d}(\Gamma+\bar{\Gamma})}{\mathrm{d}q^2}-11\right),\label{eqn:fl_counting}\displaybreak[0]
\intertext{where the latter expression makes use of the massless lepton approximation. We also study the possibility of an early measurement of $S_5$, which can be measured using only $\cos\theta_{\kstar}$ and $\phi$. It is possible to express this as}\displaybreak[0]
S_5 =& \frac{4}{3}\left(\int_{0}^{\pi/2}+\int_{3 \pi/2}^{2 \pi}-\int_{\pi/2}^{3\pi/2}\right)\mathrm{d}\phi\left( \int_0^1 - \int_{-1}^0  \right)\rm{d}\cos\theta_{\kstar} \frac{\mathrm{d}^3(\Gamma-\bar\Gamma)}{\mathrm{d}q^2\, \rm{d}\cos\theta_{\kstar} \,\rm{d}\phi}\bigg/ \frac{\mathrm{d}(\Gamma+\bar\Gamma)}{\mathrm{d}q^2}\,.\displaybreak[0]\label{eqn:s5_counting}
\end{align}
A comprehensive study of the effects of the Wilson coefficients on the above observables, and {\em vice-versa}, can be found in \Rf\cite{Altmannshofer:2008dz}. We note that $S_3$, $A_7$, and $A_9$ can also be extracted by the counting of signal events over one or two angles. \S{3} is related to the well known and theoretically clean observable \AT{2} \cite{Kruger:2005ep}; to be precise, \S{3} equals $\frac{1}{2}(1-\Fl)\AT{2}$ in the massless lepton limit. While significant enhancement of \AT{2} is possible in the presence of non-SM $\Ceffp{7}$ \cite{Lunghi:2006hc}, the $\frac{1}{2}(1-\Fl)$ prefactor implies that the enhancement is less pronounced in \S{3} \cite{Egede:1048970}. The smallness of \S{3} means that the experimental sensitivity to \rate{\S{3}} will be limited in the first few years of \lhcb data taking; thus, the study of \S{3} is thus left for other works \cite{Egede:2008uy}. Enhancements to \A{7} and \A{9} in the presence of NP phases can, however, be sizable \cite{Bobeth:2008ij} and could, in principle, lead to reasonable experimental resolutions, particularly for \rate{\A{9}}. However, these measurements will still be experimentally challenging in the first few years. For these reasons we choose to focus on \Afb, \Fl and \S{5} for early study at \lhcb.

As stated earlier, NP enters the calculations through contributions to the Wilson coefficients; constraints on these contributions are described in the \seq\ref{sec:constraints}. It is well known that for certain  values of $q^2$, the observables \Afb and $S_5$ vanish. We refer to these values of $q^2$ as the zero-crossing points, $q_0^2(\Afb)$ and $q_0^2(S_5)$. They are particularly sensitive to NP, and can be used to further constrain the values of the Wilson Coefficients. At leading order, in the large recoil limit, and for real values of the Wilson coefficients, it is possible to obtain simple expressions for $q_0^2(\Afb)$~\cite{Burdman:1998mk,Beneke:2001at} and $q_0^2(S_5)$:
\begin{align}
q_0^2(\Afb)=-2 m_B\,m_b\dfrac{ \Ceff{7}}{ \Ceff{9}};\qquad\qquad
q_0^2(S_5)= \dfrac{-m_B\,m_b(\Ceff{7}+\Ceffp{7})}{\Ceff{9}+\hat{m}_b (\Ceff{7}+\Ceffp{7})}\,.
\end{align}
In deriving these results we make use of the soft form factors, following \Refs\cite{Beneke:2001at,Beneke:2004dp}. The two observables provide complementary sensitivity to NP through their differing dependence on the Wilson coefficients, and allow for sensitivity to both chiralities of $\mathcal{O}_{7}$. The cancellation of the soft form factors and the relative smallness of $\mathcal{O}(\alpha_{s})$ corrections mean that {\em both} zero--crossing points meet the criteria for {\em theoretical cleanliness} given in, e.g., \Rf\cite{Egede:2008uy}. In addition, we define the gradient of \Afb and $S_5$ at their zero-crossing points,
\begin{align}
 G_0(\mathcal{O})=\dfrac{\mathrm{d} \mathcal{O}}{\mathrm{d}\qsq}\bigg|_{q_0^2(\mathcal{O})}\,,
\end{align}
where $\mathcal{O}$ is the observable \Afb or $S_5$ respectively. \Afb has also been studied in the context of $B\to K\pi l^+l^-$ \cite{Grinstein:2005ud}, where expressions for $q_0^2(\Afb)$ and $G_0(\Afb)$ were determined for the case of an energetic kaon and soft pion. However, the kinematic region where the $K \pi$ pair is energetic is dominated by the \Kstar, and non-resonant effects can be neglected.

\subsection{Overview of Specific Models and Effects on Wilson Coefficients}\label{sec:model-dependent}
The observables for \BdbKsmm  are most sensitive to the Wilson coefficients \Ceff{7}, \Ceff{9}, \Ceff{10} and their primed equivalents, so we concentrate on the NP contributions to these in this section. We also consider \CLR{S} and \CLR{P} for completeness; however, experimental sensitivity to their effects are expected to be limited in this decay.
\begin{itemize}
\item \textbf{Flavour Blind MSSM} (FBMSSM): Here the MFV version of the Minimal Supersymmetric Standard Model (MSSM) is modified by some flavour-conserving but \CP-violating phases in the soft supersymmetry (SUSY) breaking trilinear couplings \cite{Altmannshofer:2008hc}. The Wilson coefficients we use correspond to those calculated in scenario FBMSSM II defined in Table 11 of \Rf\cite{Altmannshofer:2008dz}. The additional \CP-violation contributes substantial complex phases to \Ceff{7}, however there is no flavour structure beyond the SM, so primed operators are suppressed as in the SM. As in all SUSY models, scalar and pseudoscalar operators arise due to the additional Higgs doublet.
\item \textbf{General MSSM} (GMSSM):  Minimal flavour violation is not imposed, and generic flavour- and \CP-violating soft SUSY-breaking terms are allowed~\cite{Gabrielli:2002me}. The Wilson coefficients we use are close to the scenario GMSSM IV in \Rf\cite{Altmannshofer:2008dz}, corresponding to large NP contributions to both \Ceff{7} and \Ceffp{7} allowed by existing experimental bounds (see \seq\ref{sec:constraints}).
\end{itemize}

\begin{figure}
\centering
      \includegraphics[width=0.48\textwidth]{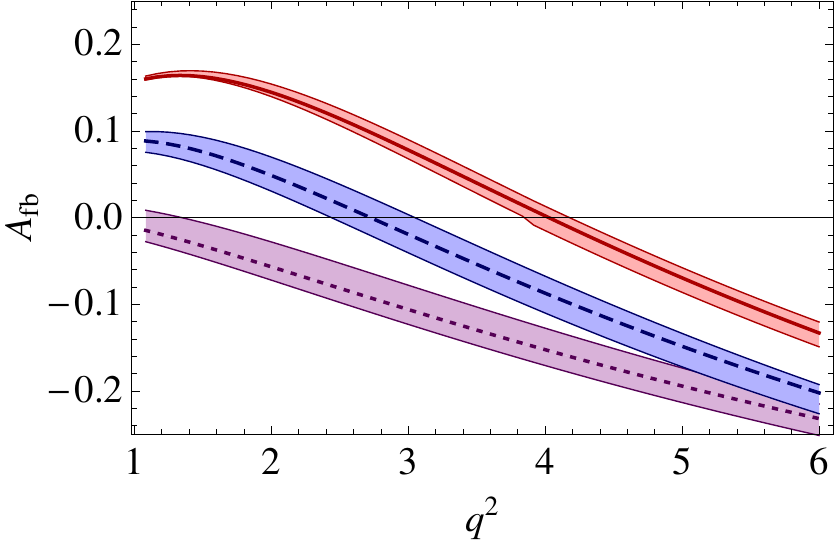} 
      \includegraphics[width=0.47\textwidth]{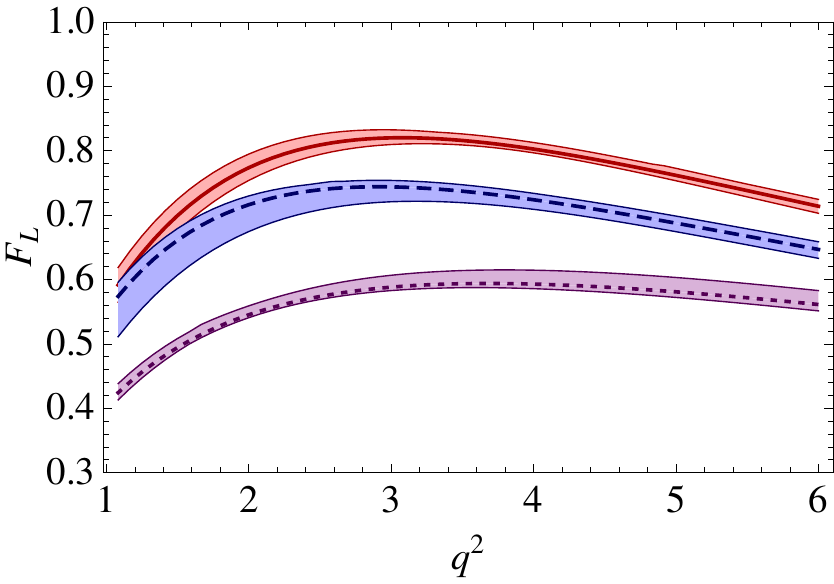} 
      \includegraphics[width=0.48\textwidth]{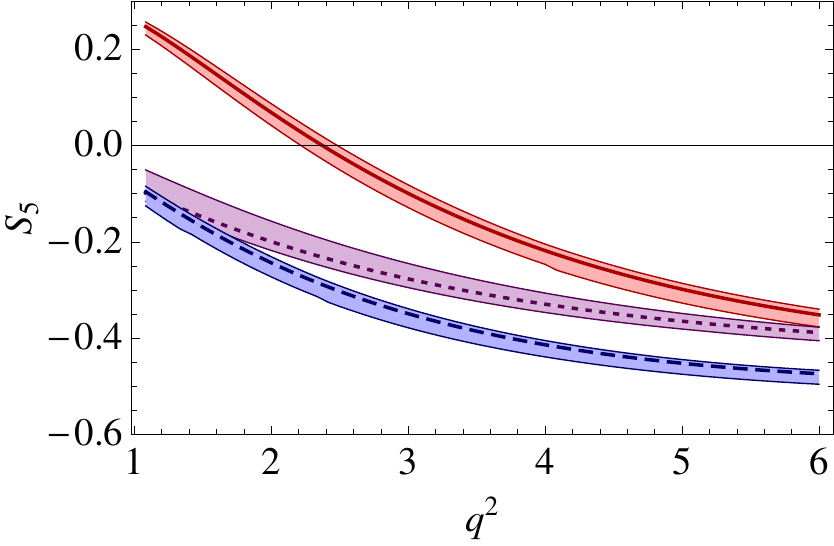} 
\caption{Theoretical predictions for $A_{\mathrm{FB}}$, $F_L$ and $S_5$. The red (continuous) line is the SM, the blue (dashed) line is the GMSSM, and the purple (dotted) line is the FBMSSM.}
\label{fig:TheoryPlots}
\end{figure}

The Wilson coefficients in the above scenarios are given explicitly in \tab\ref{tab:wilson_mb}. The central values for the distributions of \Afb, \Fl, and \S{5} are shown in \fig\ref{fig:TheoryPlots} for the SM, the GMSSM, and FBMSSM, along with estimates of the theoretical uncertainties. The agreement with previous results is good. The predominant sources of the uncertainties are the form factors, hadronic parameters, and quark masses, which are determined as discussed in \seq\ref{sec:theory}. We also include the uncertainty arising from varying the factorization scale, $\mu$, in the range $\mu \in [\mu/2,2\mu]$. The three distributions all show significant variation for the models considered here, as do the position or absence of the zero-crossing points in \Afb and \S{5} in the range $\qsq \in [1,6]\,\mathrm{GeV}^2$.

\begin{table}
\setlength{\extrarowheight}{0.3em}
\centering
\begin{tabular}{| c |c|c|c|}
\hline
\T &\multicolumn{3}{c|}{Model} \\
\hline
\T& SM & FBMSSM & GMSSM \\
\hline
\hline
\T $\Ceff{7}(\mu)$   & -0.306  & \phantom{-}0.031+0.475i  & -0.186+0.002i \\
$\Ceffp{7}(\mu)$   & -0.007  & \phantom{-}0.008+0.003i  & \phantom{-}0.155+0.160i  \\
$\Ceff{8}(\mu)$  &  -0.159 & -0.085+0.149i & -0.062+0.004i  \\
$\Ceff{8}(\mu)$   & -0.004  & -0.000+0.001i & \phantom{-}0.330+0.336i  \\
$\Delta\Ceff{9}(\mu)$   & 4.220   & \phantom{-}4.257+0.000i        & \phantom{-}4.231+0.000i \\
$\Ceffp{9}(\mu)$   &  0.000       & \phantom{-}0.002+0.000i       & \phantom{-}0.018+0.000i     \\
$\Ceff{10}(\mu)$  & -4.093  & -4.063+0.000i  & -4.241+0.000i \\
$\Ceffp{10}(\mu)$  &  0.000       & \phantom{-}0.004+0.000i    & \phantom{-}0.003+0.003i      \\
$\hat{C}_{S}(\mu)/\mathrm{GeV}^{-1}$ &   0.000      & -0.044-0.056i & \phantom{-}0.000+0.001i      \\
\Bo$\hat{C}_{P}(\mu)/\mathrm{GeV}^{-1}$ &   0.000      & \phantom{-}0.043+0.054i  & \phantom{-}0.001+0.001i       \\
\hline
\end{tabular}
\caption{\label{tab:wilson_mb} NP Wilson coefficients at $\mu = m_{\b,\mathrm{PS}}(2\gevcc) = 4.52\gevcc$ in the FBMSSM and GMSSM as described in \seq\ref{sec:model-dependent}, where $\hat{C}_{X}(\mu)=(\CL{X}-\CR{X})(\mu)$ for $X=S$ or $P$.}
\end{table}

\section{Constraints}\label{sec:constraints}
Experimental results can be used to constrain the NP contributions, denoted $C_i^\mathrm{NP}$, to the Wilson coefficients: we define $C_i=C_i^\mathrm{SM}+C_i^\mathrm{NP}$. We can then determine possible model-independent effects of NP on \BdbKsmm. The most important constraints on the Wilson coefficients are from the following measurements:
\begin{itemize}
\item \textbf{Branching Ratio for $B_s\to\mu^+\mu^-$}: This is used to constrain the possible NP contribution to the scalar and pseudoscalar operators. To calculate the branching ratio we use the standard result from \Rf\cite{Altmannshofer:2008dz}
\begin{equation}
\mathcal{B}(B_s\to\mu^+\mu^-)=\tau_{B_s}f_{B_s}^2 m_{B_s}\frac{\alpha_{EM}^2 G_F^2}{16 \pi ^3} |V_{tb}V^*_{ts}|^2\sqrt{1 - \frac{4 m_\mu^2}{m_{B_s}^2}}(|S|^2 \left(1 - \frac{4 m_\mu^2}{m_{B_s}^2}\right) +|P|^2),
\end{equation}
with the definitions 
\begin{equation}
S=\frac{m_{B_s}^2}{2}(\CL{S}-\CR{S}); \qquad P=\frac{m_{B_s}^2}{2}(\CL{P}-\CR{P})+m_\mu(\Ceff{10}-\Ceffp{10}).
\end{equation}
We use $f_{B_s}=0.259\pm0.032$ GeV \cite{Gray:2005ad}, $\tau_{B_s}=1.456\pm0.03$ps \cite{Barberio:2007cr} and $m_{B_s}=5.37$ GeV \cite{Amsler:2008zz}, and other numerical parameters as in \Rf\cite{Altmannshofer:2008dz}. In agreement with existing results, we find the SM prediction, $\mathrm{BR}(B_s\to\mu^+\mu^-)=(3.70\pm 0.31)\cdot10^{-9}$, to be well below the current experimental upper bound $3.6 \cdot 10^{-8}$ \cite{CDF:2007kv,*CDF:2009}.
 
\item \textbf{Branching Ratio for $B\to X_s l^+ l^-$}: We compare NP predictions for $\mathcal{B}(B\to X_s l^+l^-)_{\text{1-6}\,\mathrm{GeV}^2}$ to the mean experimental value $(1.60 \pm 0.51)\cdot10^{-6}$, as adopted in \Rf\cite{Bobeth:2008ij}, combining the results of \babar, $(1.8 \pm 0.7\pm 0.5)\cdot10^{-6}$~\cite{Aubert:2004it}, and \belle, $(1.49_{-0.32}^{+0.41}\pm 0.50)\cdot10^{-6}$~\cite{Iwasaki:2005sy}.  This helps to constrain the NP contribution to \Ceffpn{7,9,10} as well as $C^{(')}_{S,P}$. As an inclusive mode, the calculation for the region $\qsq \in [1,6]\gev^2$ of the branching ratio is theoretically clean. We use the expression for the differential decay distribution in \Rf\cite{Hiller:2003js}, but also include the NLO corrections computed in \Rf\cite{Ghinculov:2003qd}, and the contribution of the primed operators as in \Rf\cite{Guetta:1997fw}. Using our parameters we find $\mathcal{B}(B\to X_s l^+l^-)=(1.96\pm0.11)\cdot10^{-6}$ for the SM.

\item \textbf{Branching Ratio for $B\to X_s \gamma $}: The current experimental average for $\mathrm{E}_\gamma > 1.6\,\mathrm{GeV}$ is $\,\mathcal{B}(B\to X_s\gamma)=(3.52\pm 0.23\pm 0.09 )\cdot 10^{-4}$, as calculated by the Heavy Flavor Averaging Group \cite{Barberio:2007cr}. We use the recent theoretical SM result of \Rf\cite{Gambino:2008fj}, $(3.28\pm 0.25)\cdot10^{-4}$ for $\mathrm{E}_\gamma > 1.6\,\mathrm{GeV}$, and include NP effects as in \Rf\cite{Lunghi:2006hc}. The SM calculation makes use of the kinetic renormalization scheme for determining $m_c$ and \(m_{\b}\); an alternative calculation using the 1S scheme leads to a branching ratio of $(3.15\pm 0.23)\cdot10^{-4}$ \cite{Misiak:2006zs,Misiak:2008ss}; however, our results are not sensitive to the difference between these two values.

\item \textbf{Time dependent \CP Asymmetry $S(B\to K^* \gamma$)}: This constraint  is sensitive to the photon polarization, and, hence, to \Ceffp{7}. This should be compared to $S(B\to K^* \gamma)=(-1.6\pm 2.2)\cdot 10^{-1}$ from experiment \cite{Barberio:2007cr}. Our SM result $S(B\to K^* \gamma)=(-0.26\pm0.05 )\cdot10^{-1}$ agrees with that of \Rf\cite{Bobeth:2008ij} within uncertainties. In \Refs\cite{Ball:2006cva,Ball:2006eu}, the soft gluon contribution was calculated, leading to a small correction to the predicted value. This is neglected in our treatment as it has little effect on the constraining power of the experimental measurement.

\item \textbf{Integrated Forward-Backward Asymmetry $\langle \Afb\rangle_{\text{1-6}\,\mathrm{GeV}^2}$ for \BdbKsmm}: We use the existing \BdbKsmm measurements as constraints. Recently \belle has made a measurement of the forward-backward asymmetry, and finds the integrated $A_\mathrm{FB}$ value in the region 1-6$\,\mathrm{GeV}^2$ to be $-0.26\pm0.29$ \cite{Belle:2009zv}. This is to be compared to our SM prediction of $0.04\pm0.03$, which is in agreement with the recent result in \Rf\cite{Bauer:2009cf}. This observable constrains the Wilson coefficients as seen in \tab\ref{tab:OsandWCs}. We look forward to a 1-6$\,\mathrm{GeV}^2$ measurement from CDF with great interest \cite{cdf_hcp2009_ksmm}.

\item \textbf{Integrated Longitudinal Polarization Fraction $\langle \Fl\rangle_{\text{1-6}\,\mathrm{GeV}^2}$ for \BdbKsmm}: \belle has also recently measured the Longitudinal Polarization Fraction to be $0.67\pm0.24$ \cite{Belle:2009zv}. This should be compared to our SM prediction $0.76\pm0.08$, also in agreement with \Rf\cite{Bauer:2009cf}. Again this constraint affects Wilson coefficients as seen in \tab\ref{tab:OsandWCs}.

\end{itemize}

\begin{table}
\centering
\setlength{\extrarowheight}{0.3em}
\begin{tabular}{|c|c|}
\hline
\T\,Observable\, & Wilson Coefficients\\
\hline
\hline
\T \Afb & \Ceff{7}, \Ceff{9}\\
\Fl & \,\Ceff{7}, \Ceffp{7},\Ceff{8},\Ceffp{9}, \Ceff{10}, \Ceffp{10}\,\\
\Bo \S{5} & \Ceff{7}, \Ceffp{7}, \Ceff{9}, \Ceffp{10}\\
\hline
\end{tabular}
\caption{Relevant observables and the Wilson coefficients they most strongly depend on \cite{Altmannshofer:2008dz}.}
\label{tab:OsandWCs}
\end{table}

\begin{table}
\setlength{\extrarowheight}{0.3em}
\centering
\begin{tabular}{|c|c|c|}
\hline
\T Observable & Experiment & SM Theory\\
\hline
\hline
\T$ \mathcal{B}(B_s\to\mu^+\mu^-)$ &  $3.6 \cdot 10^{-8}$~\cite{CDF:2007kv,*CDF:2009} & $(3.70\pm 0.31)\cdot10^{-9}$\\

$\,\mathcal{B}(B\to X_s l^+l^-)_{\text{1-6}\,\mathrm{GeV}^2}\,$ & $(1.60 \pm 0.51)\cdot10^{-6}$~\cite{Bobeth:2008ij} & $(1.97\pm0.11)\cdot10^{-6}$\\

$\mathcal{B}(B\to X_s\gamma)$&\, $(3.52\pm 0.23\pm 0.09 )\cdot 10^{-4}$~\cite{Barberio:2007cr}\,& $(3.28\pm 0.25)\cdot10^{-4}$\\

 $S(B\to K^* \gamma)$& $(-1.6\pm 2.2)\cdot 10^{-1}$~\cite{Barberio:2007cr}& $(-0.26\pm0.05 )\cdot10^{-1}$\\

$\langle A_{\mathrm{FB}}\rangle_{\text{1-6}\,\mathrm{GeV}^2}$ & $-0.26\pm0.29$~\cite{Belle:2009zv}& $0.04\pm0.03$\\

\Bo$\langle F_L\rangle_{\text{1-6}\,\mathrm{GeV}^2}$ & $0.67\pm0.24$~\cite{Belle:2009zv} & $0.76\pm0.08$\\
\hline
\end{tabular}
\caption{\label{tab:constraints}Experimental measurements used as constraints, along with theoretical predictions in the SM.}
\end{table}

In order to assess the impact of these constraints on the NP contributions to the underlying Wilson coefficients in as general a way as possible, we have performed a semi-random walk through parameter space. We allow $(\CL{S}-\CR{S})$, $(\CL{P}-\CR{P})$ and the NP components of \Ceffpn{7-10} to vary simultaneously, both in magnitude and phase. To our knowledge this has not been done in previous studies. At each randomly chosen point in parameter space, predictions are made for the six observables listed above. The point is then either accepted or rejected using a modified $\chi^2$ metric that treats experimental uncertainties as being normally distributed, but theoretical uncertainties as having uniform probability within the specified range. Following traditional minimization techniques, the random walk is guided by this modified $\chi^2$ so that regions with lower values may be identified. Using this method, a sample of $2.5\cdot 10^5$ independent sets of Wilson coefficients was produced. Each set results in predictions for the observables listed above with better than $2\sigma$ agreement with current measurements. It was found that the agreement between existing measurements and the SM is excellent, with a $\chi^2$ per degree of freedom of $0.35$. While this is not implausible for six degrees of freedom, the level of agreement suggests that more detailed study of the theoretical uncertainties will be required as experimental resolutions improve.

\begin{figure}
\centering
\includegraphics[width=0.45\textwidth]{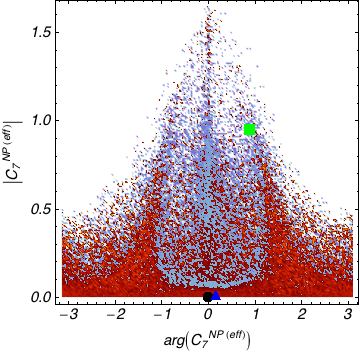}   \includegraphics[width=0.45\textwidth]{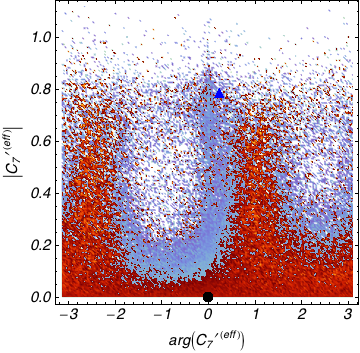}
\caption{\label{fig:constraint-2009-C7}[Colour online] Allowed parameter space for the NP contribution to \Ceff{7} and \Ceffp{7} at the scale $\mu = m_{W}$, as described in \seq\ref{sec:constraints}. Points with a compatibility with data of 68\% or better are drawn with a dark (red) colour palette, while those with less than this are drawn with a light (blue) palette. The SM point is shown in black at the origin, while the FBMSSM is a green square and the GMSSM is a blue triangle. The Wilson coefficients for these models are shown in \tab\ref{tab:wilson_mb}.}
\end{figure}

\fig\ref{fig:constraint-2009-C7} shows the range of values found for the phase and magnitude of the NP contribution to \Ceff{7} and \Ceffp{7} (at the scale $\mu = m_{\W}$) during the parameter space exploration. The colour index shows the mean value of the probability that a point is compatible with current experimental results. Areas with probability greater than $1\sigma$ are shaded red, while those with less than $1\sigma$ are shaded blue. The outline of the $1\sigma$ contour can clearly be seen. The values of the Wilson coefficients for the SM, FBMSSM, and GMSSM are also shown. 

\fig\ref{fig:constraint-2009-C7} can be compared to \fig2 from \Rf\cite{Bobeth:2008ij}, in which \Ceff{7} and \Ceffp{7} are assumed to be real and all other Wilson coefficients SM-like. The effects of weakening these assumptions can be seen. Similar figures are shown for the other Wilson coefficients in \figs\ref{fig:constraint-2009-all} and \ref{fig:constraint-2009-all2}. The allowed regions of parameter space are still large, particularly if NP phases are allowed. In contrast to \Rf\cite{Bobeth:2008ij}, constraints from \Afb measurements at high--\qsq (low recoil) are not included as we feel that NLO effects are not under control in this region. The effect of this constraint may be seen by comparing our \Ceff{10} figure, shown in \fig\ref{fig:constraint-2009-all2}, with that in \fig2 of \Rf\cite{Bobeth:2008ij}.

\begin{figure}
\centering
      \includegraphics[width=0.45\textwidth]{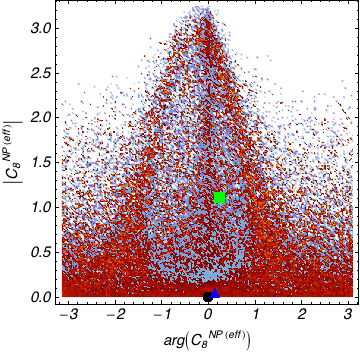}
      \includegraphics[width=0.45\textwidth]{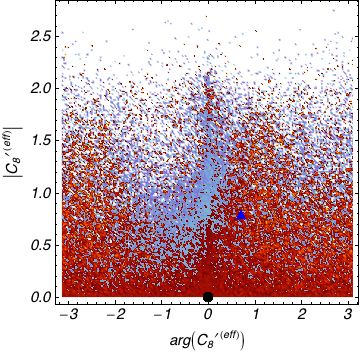}
      \includegraphics[width=0.45\textwidth]{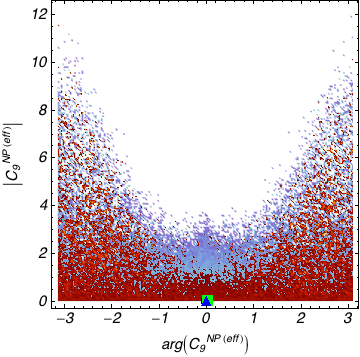}
      \includegraphics[width=0.45\textwidth]{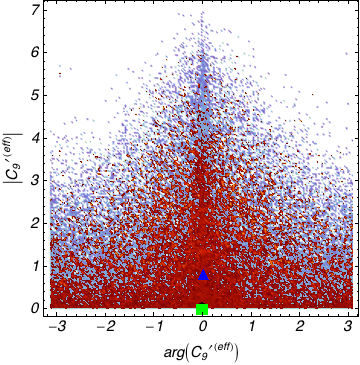}
    \caption{\label{fig:constraint-2009-all}[Colour online] Allowed parameter space for the Wilson coefficients \Ceffpn{8-9} after applying relevant $\b\to s$ experimental constraints. The colour coding is the same as in \fig\ref{fig:constraint-2009-C7}.}
\end{figure}

\begin{figure}
\centering
      \includegraphics[width=0.45\textwidth]{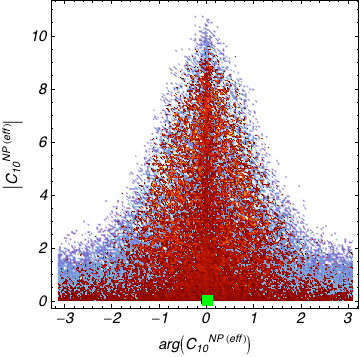}
      \includegraphics[width=0.44\textwidth]{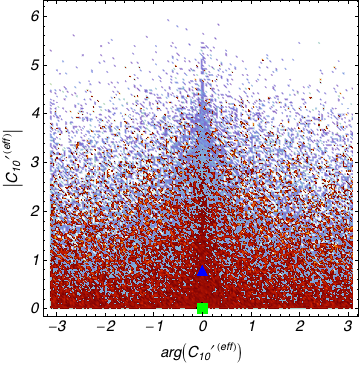}
      \includegraphics[width=0.45\textwidth]{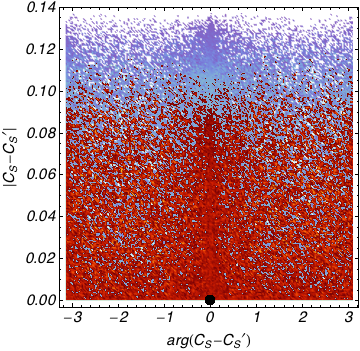}
      \includegraphics[width=0.445\textwidth]{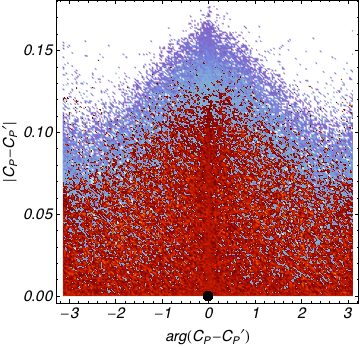}
\caption{\label{fig:constraint-2009-all2}[Colour online] Allowed parameter space for the Wilson coefficients \Ceffpn{10} and $(\CL{S,P}-\CR{S,P})$ after applying relevant $\b\to s$ experimental constraints. The colour coding is the same as in \fig\ref{fig:constraint-2009-C7}.}
\end{figure}

The ensemble of constrained NP models can also be used to explore the likely values of the \Afb and \S{5} zero-crossing points in the range $\qsq \in [0.5,15]\gev^2$. While it should be noted that theoretical uncertainties are not well controlled over this \qsq range, the majority of points within the $1\sigma$ contour lie within the theoretically clean region, $\qsq \in [1,6]\gev^2$ (see  \fig\ref{fig:constraints-zero}). It was found that 8\% of the parameter space points considered had no \Afb zero-crossing in the range $\qsq \in [0.5,15]\gev^2$. For \S{5}, only 2\% of points had no zero-crossing in the same range. \fig\ref{fig:constraints-grad} shows the \Afb and \S{5} gradients at their zero-crossing points. We find that, for the majority of points, $G_0(\S{5})$ is greater than $G_0(\Afb)$. This will have an impact for the $q^{2}_{0}(\S{5})$ experimental analysis discussed in the \seq\ref{sec:zero-crossing}.

To summarize, in this section we have considered six existing experimental constraints, and used these to determine the allowed regions in parameter space for the NP contribution to the Wilson coefficients. These allowed values for the Wilson coefficients were then used to find corresponding predictions for $q^{2}_{0}(\S{5})$, $q^{2}_{0}(\Afb)$, $G_0(\S{5})$, and $G_0(\Afb)$. In the following sections, we investigate the experimental sensitivity to the observables \Afb, \S{5}, and \Fl, and how measurements of these could have an impact on the allowed NP contributions to the Wilson coefficients.

\begin{figure}
\centering
    \subfloat[]{
      \label{fig:constraints-zero}
      \includegraphics[width=0.43\textwidth]{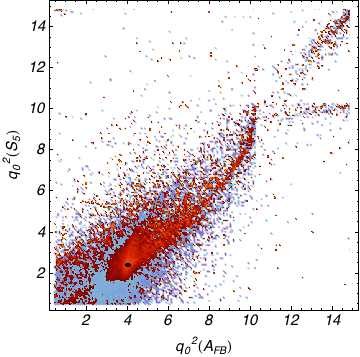}
    }
    \subfloat[]{
      \label{fig:constraints-grad}
      \includegraphics[width=0.45\textwidth]{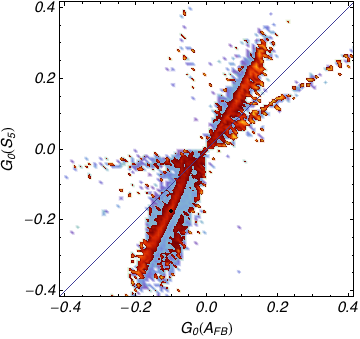}
    }
\caption{\label{fig:constraints_s56_zeros}[Colour online] \fig\subref{fig:constraints-zero} shows allowed values of the \Afb and \S{5} zero-crossing points in the range $\qsq \in [0.5,15]\gev^2$. The SM point and its uncertainty is shown as a black ellipse. \fig\subref{fig:constraints-grad} shows the gradient of the \Afb and \S{5} at the zero-point. For comparison, the line $G_{0}(\S{5}) = G_{0}(\Afb)$ is included. In each case the colour index has the same meaning as in \fig\ref{fig:constraint-2009-C7}.}
\end{figure}

\section{Experimental Sensitivities}\label{sec:sense}
Three observables that can be measured as a function of $q^2$ by counting signal events in specific angular bins, using \eqs(\ref{eqn:afb_counting})--(\ref{eqn:s5_counting}), were highlighted in \seq\ref{sec:observables}: \Afb, \Fl, and \S{5}. These observables should be suitable for early measurement at \lhcb. In the following, we estimate the experimental sensitivities in order to make a fair comparison between these observables.

\lhcb is expected to collect $\sim 6.2\cdot 10^3$ signal events per 2\invfb of integrated luminosity with a signal to background ratio of approximately four \cite{LHCb:2009ny,Patel:1157434}. With relatively small data sets it should be possible to extract the values of these observables integrated over \qsq. These measurements provide an early opportunity to discover NP in $\b\to\s$ transitions. For larger data sets it will be possible to map out the dependence on \qsq as well, allowing for additional NP discrimination. Studies of these two approaches can be found in \Refs\cite{Dickens:1045395,Jansen:1156131} for the observable \Afb.

To assess the impact of each potential measurement on the allowed NP parameter space, simple analyses have been developed to extract the \qsq integrated values of \Afb, \Fl, and \S{5} in the region $\qsq \in [1,6]\gev^2$. In addition, analyses have been constructed to extract the \qsq dependence of \Afb and \S{5}, along with their zero-crossing points; the latter can be found numerically from the $\Afb(\qsq)$ and $\S{5}(\qsq)$ distributions. In order to minimize the experimental uncertainties on these points, a larger region of $\qsq \in [0.5,8.5]\gev^2$ was used for these analyses following \Rf\cite{Jansen:1156131}. An ensemble of 1200 simulated \BdbKsmmFull data sets was created, each containing the (Poisson fluctuated) number of signal and background expected from 2\invfb of integrated luminosity at \lhcb. Other integrated luminosities were obtained by linearly scaling the yield estimates.  Each analysis was then run in turn on the data sets in order to estimate the statistical uncertainty expected for each measurement. This allows for a fair comparison to be made between observables for a given integrated luminosity.

\subsection{Data Set Generation}\label{sec:generation}
The theoretical framework introduced in \seq\ref{sec:theory} was implemented as a plug-in for the standard decay tree simulation tool \evtgen \cite{Lange:2001uf}. This  allows \BdbKsmmFull events to be simulated. A simplified background sample was generated separately. This was flat in the three decay angles defined in \seq\ref{sec:observables} but followed the signal distribution in \qsq and a gently falling exponential in the \B invariant mass, $m_{\B}$. All events had $m_{\B}$ within a wide window \(^{+250}_{-150}\mev\) around the nominal \B mass. A central signal region was also defined with width $\pm 50\mev$. Events outside of this region were assumed to be part of a background dominated side-band. Signal and background events were generated following the relative normalization given in \Refs\cite{LHCb:2009ny,Patel:1157434}. For each event in a data set, the three decay angles, \qsq and $m_{\B}$ were determined and used as input for each analysis.

\subsection{\qsq Integrated Analyses}\label{sec:integrated}
The integrated quantities can be extracted by estimating the number of signal events in each angular bin using a fit to the $m_{\B}$ distribution. The signal contribution was parametrized as a Gaussian with an exponential tail, while the background was modelled as an exponential with a negative coefficient. A fit was performed to each data set to extract the signal and background shape parameters for that sample. Each sample was then reduced into the relevant angular bins. For \Afb, following \eq(\ref{eqn:afb_counting}) these bins would be $\cos\thetaL \in [-1,0]$ and $\cos\thetaL \in [0,1]$ for all events in the range $\qsq \in [1,6]\gev^2$. To extract an estimate of the number of signal and background events in each angular bin, a separate fit to the $m_{\B}$ signal and background distributions was then performed, keeping all shape parameters fixed. The value of \rate{\Afb} was determined with \eq(\ref{eqn:afb_counting}). A similar procedure was applied to \eqs(\ref{eqn:fl_counting}) and (\ref{eqn:s5_counting}) to extract \rate{\Fl} and \rate{\S{5}}.
\vspace{2cm}

\subsection{\qsq Dependent Analyses}\label{sec:dependent}
Following \Rf\cite{Jansen:1156131}, a polynomial shape was fit to the \qsq distribution in each angular bin. The method proceeds as in \seq\ref{sec:integrated}, using the \B mass distribution to find the total number of signal and background events in each angular bin. However, the background shape extracted is used to estimate the number of signal events in the \B mass signal window. The \qsq dependence of the signal and background distributions was parametrized using second and third order Chebyshev polynomials respectively. A simultaneous fit in the signal and side-band regions of the \B mass distribution was used to determine the shape parameters of signal and background polynomials using the relative signal/background normalization found from the \B mass fits. In the case of \Afb, the procedure would lead to the extraction of two \qsq dependent signal polynomials: one for events with $\cos\thetaL \in [-1,0]$ and the other for $\cos\thetaL \in [0,1]$. The value of \Afb(\qsq) can then be found using these polynomials and \eq(\ref{eqn:afb_counting}). The \Afb zero-crossing point was found numerically from the combined functions. A similar approach was applied to \S{5} and its zero-crossing; however, six angular bins in \thetaK and $\phi$ were required.

\subsection{Results}\label{sec:sense-results}
When comparing different observables and analyses it is useful to consider the mean expected experimental sensitivity for a given integrated luminosity. These expected sensitivities can be calculated from the ensemble of toy \lhcb experiments introduced in \seqs\ref{sec:integrated} and \ref{sec:dependent}. 1200 individual experiments were performed, and for each one a value of, for example, $\qsq_{0}(\Afb)$ was found. Following \Rf\cite{Egede:2008uy}, the mean, one and two sigma contours could then be found from these results. The method used allows for non-normally distributed results by putting the ensemble in numerical order and then selecting the values closest to the contour\footnote{For the one sigma bound these would be the \(188^{\mathrm{th}}\) and  \(1010^{\mathrm{th}}\) results in the ordered ensemble.}. Any biases introduced can be identified by comparing the median result and input value. Example ensembles are shown in \fig\ref{fig:unbinned-2fb-zero} for $\qsq_{0}(\Afb)$ and $\qsq_{0}(\S{5})$, assuming 2\invfb of \lhcb data and following the SM.
\subsubsection{Integrated Quantities}
The estimated $1\sigma$ sensitivities for the integrated observables \rate{\Afb}, \rate{\Fl} and \rate{\S{5}} for toy \lhcb data set sizes of 2\invfb, 1\invfb and 0.5\invfb are shown in \tab\ref{tab:sense}. Any differences between the input and extracted median values were seen to be small relative to the estimated uncertainties. The estimated \lhcb experimental uncertainties are of a similar size to the current theoretical uncertainties, and much smaller than the current experimental constraints \cite{Belle:2009zv}.

\begin{table}
\setlength{\extrarowheight}{0.6em}
\begin{center}
\begin{tabular}{| l |c|c|c|}
\hline
\T\,\,Observable & \,2\invfb\, & \,1\invfb \,&\, 0.5\invfb\, \\
\hline
\hline
\,$\rate{\Afb}$ & $^{~+0.03}_{~-0.04}$ & $^{~+0.05}_{~-0.03}$ & $^{~+0.08}_{~-0.06}$\\
\,$\rate{\Fl}$ & $^{~+0.02}_{~-0.02}$ & $^{~+0.04}_{~-0.03}$ &$^{~+0.04}_{~-0.06}$\\
\,$\rate{\S{5}}$ & $^{~+0.07}_{~-0.08}$ & $^{~+0.09}_{~-0.11}$ & $^{~+0.16}_{~-0.15}$\\
$q^{2}_{0}(\Afb)$ & $^{~+0.56}_{~-0.94}$ & $^{~+1.27}_{~-0.97}$ & -- \\
\Bo\,$q^{2}_{0}(\S{5})$ & $^{~+0.27}_{~-0.25}$ & $^{~+0.53}_{~-0.40}$ & -- \\
\hline
\end{tabular}
\end{center}
\caption{\label{tab:sense} Estimated $1\sigma$ \lhcb sensitivities for 2\invfb, 1\invfb and 0.5\invfb of integrated luminosity, assuming the SM.}
\end{table}%

\subsubsection{Zero-Crossings}\label{sec:zero-crossing}
\begin{figure}
\centering
      \includegraphics[width=0.48\textwidth]{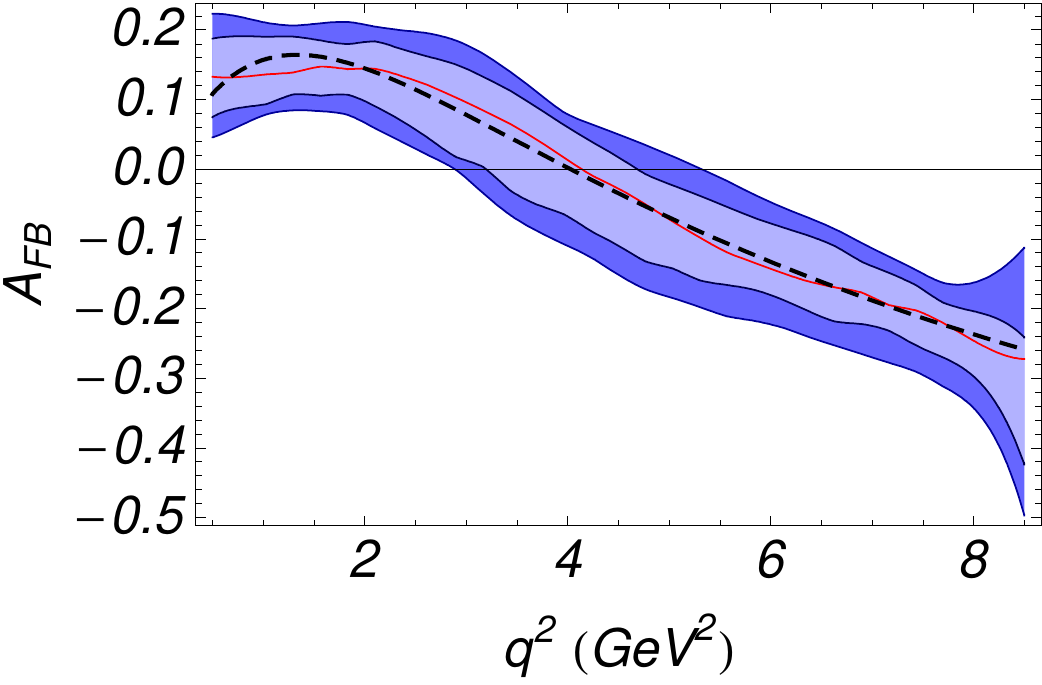}
      \includegraphics[width=0.48\textwidth]{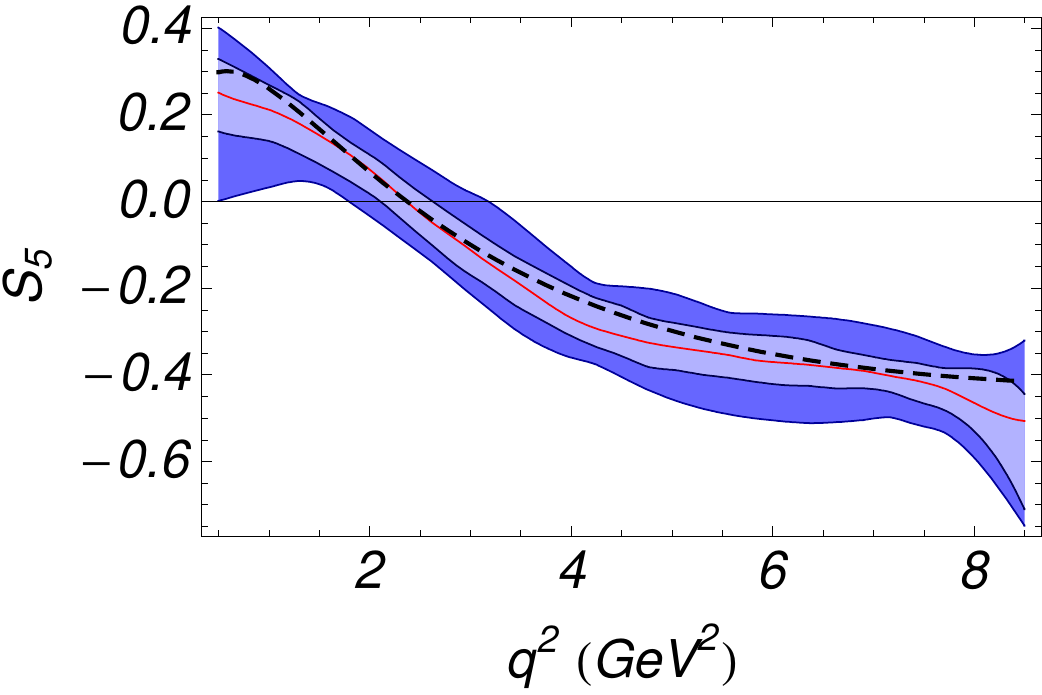}
\caption{\label{fig:unbinned-2fb}Projected experimental sensitivities to the observables \Afb and \S{5} using an unbinned polynomial fit to 2\invfb of \lhcb data in the range $\qsq \in [0.5,8.5]\gev^2$. The dashed line shows the input distribution, while the solid line shows the median of an ensemble of 1200 fits. The light and dark contours show the estimated one and two $\sigma$ contours.}
\end{figure}

\begin{figure}
\centering
      \includegraphics[width=0.48\textwidth]{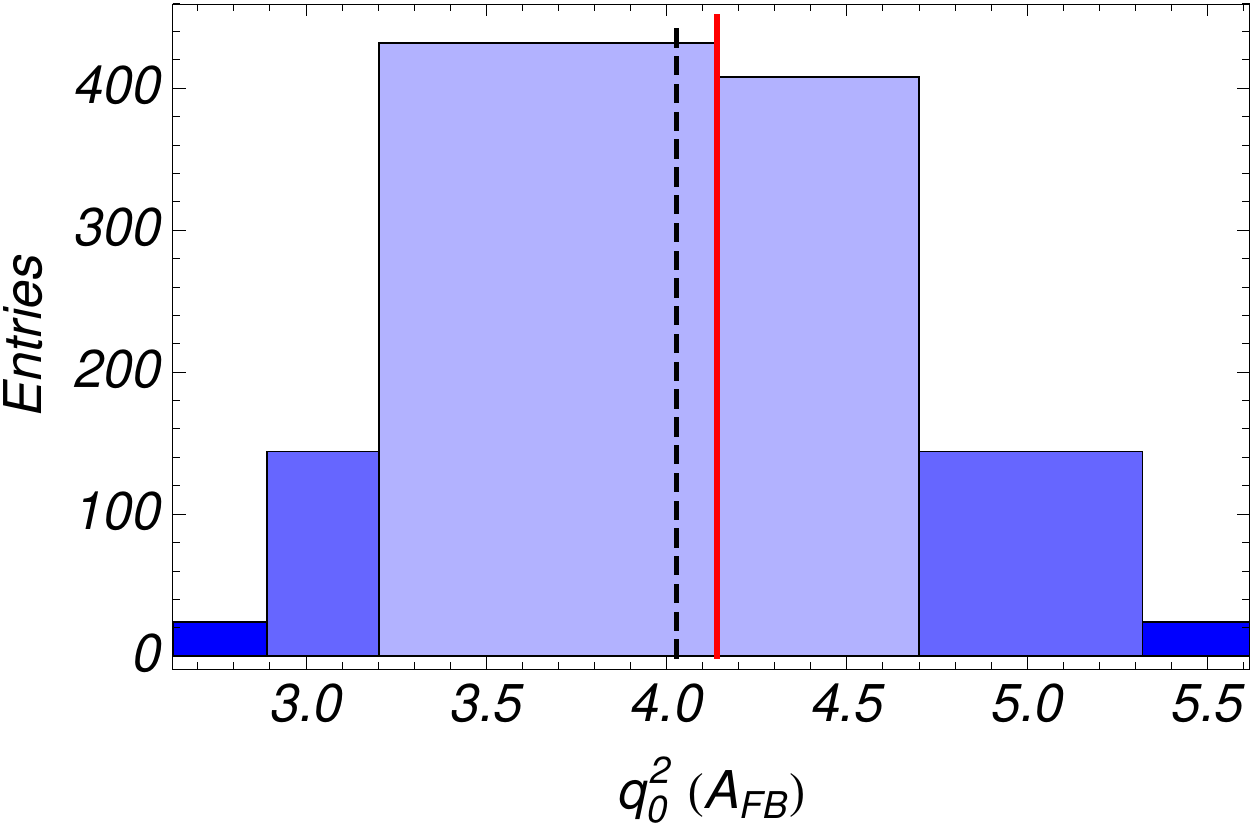}
      \includegraphics[width=0.48\textwidth]{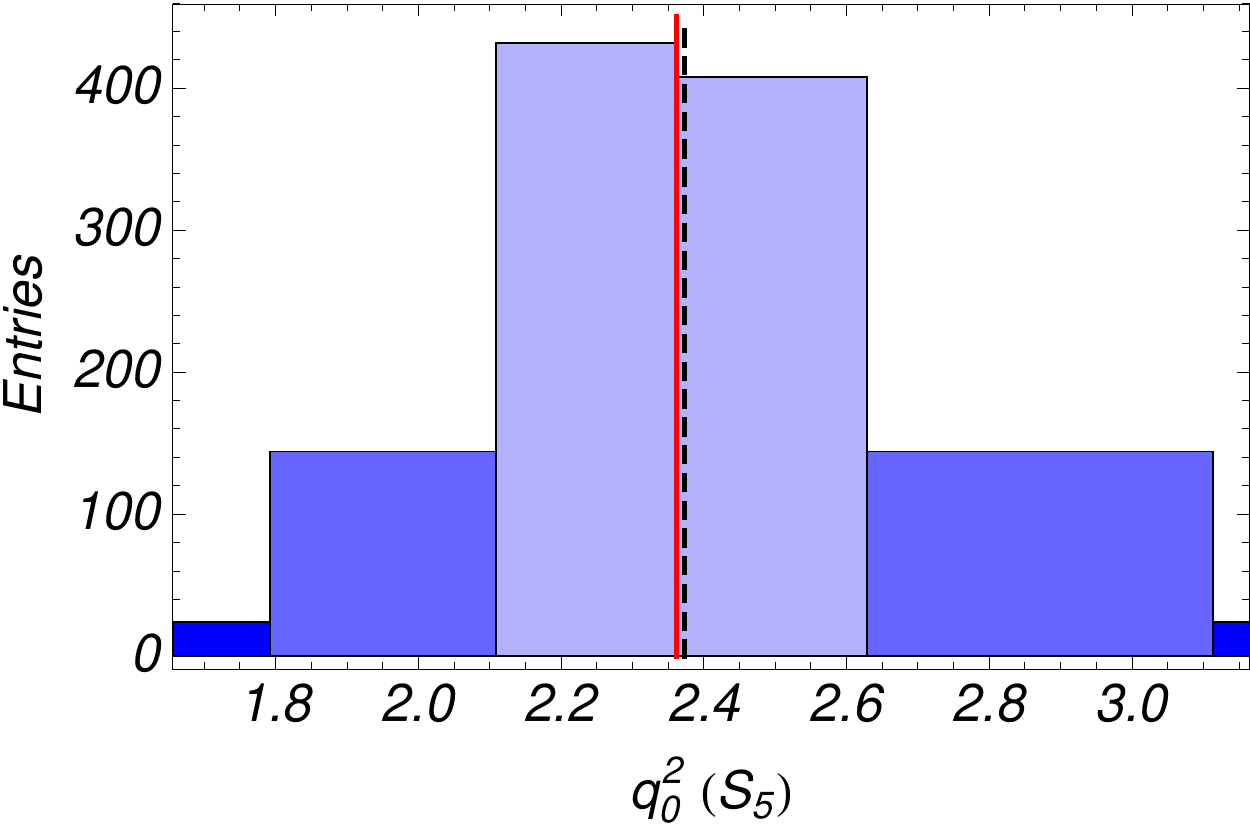}
\caption{\label{fig:unbinned-2fb-zero}Projected experimental sensitivities to the zero-crossings of \Afb and \S{5} using an unbinned polynomial fit to 2\invfb of \lhcb data in the range $\qsq \in [0.5,8.5]\gev^2$. The colour coding is the same as in \fig\ref{fig:unbinned-2fb}.}
\end{figure}
\fig\ref{fig:unbinned-2fb} shows the projected experimental sensitivity to the full \Afb and \S{5} distributions for 2\invfb of \lhcb SM data. For ease of comparison with SM predictions, the zero-crossing point is extracted from the \qsq dependent distributions. These are shown in \fig\ref{fig:unbinned-2fb-zero} for the same data sets as used in \fig\ref{fig:unbinned-2fb}. The estimated $1\sigma$ uncertainties are shown in \tab\ref{tab:sense}. As discussed in \Rf\cite{Egede:2008}, the experimental uncertainty will scale approximately linearly with the gradient at the zero-crossing, leading to the large difference in estimated sensitivities seen for $q^{2}_{0}(\Afb)$ and $q^{2}_{0}(\S{5})$ in \tab\ref{tab:sense}.

The difference in gradients between \Afb and \S{5}, seen in \fig\ref{fig:constraints-grad} for the majority of NP points, makes $q^{2}_{0}(\S{5})$ an attractive experimental target, assuming that any practical difficulties associated with the \thetaK and $\phi$ decay angles can be overcome. We see that the relative steepness of the \S{5} distribution is such that the experimental uncertainty on $q^{2}_{0}(\S{5})$ should be competitive with that on $q^{2}_{0}(\Afb)$ for the majority of the allowed regions of parameter space. For 0.5\invfb, biases on the zero-crossing points become significant when using the unbinned analysis technique; however, it is likely that coarse estimates of $q^{2}_{0}(\Afb)$ and $q^{2}_{0}(\S{5})$ could be extracted even at this relatively small integrated luminosity using alternative techniques, such as those discussed in \Rf\cite{Dickens:1045395}.  

\section{Impact of Future Measurements}\label{sec:impact}
The relative impact of the different analyses presented in \seq\ref{sec:sense} can be assessed by revisiting the parameter space exploration performed in \seq\ref{sec:constraints}. We are interested in how including these new measurements would affect the current constraints on parameter space. It is assumed that \lhcb will make 2\invfb measurements of the observables \rate{\Afb}, \rate{\S{5}}, \rate{\Fl}, $q^{2}_{0}(\Afb)$, and  $q^{2}_{0}(\S{5})$ and that the resulting experimental uncertainties are symmetrized versions of those given in \tab\ref{tab:sense}.  In addition, we assume that the measured values of these observables are not affected by NP, and are as given in \tab\ref{tab:constraints}. The total $\chi^2$ for each point in parameter space is then updated to reflect these hypothetical SM measurements. Where individual measurements are superseded by \lhcb measurements, they are replaced with no attempt at combination. However, other constraints, such as $\BR(B\to X_s\gamma)$, are included as before. In this way the constraining power of each analysis can be compared.

\begin{figure}
\centering
    \subfloat[~\rate{\Afb} \&  $q^{2}_{0}(\Afb)$]{
      \label{fig:constraints-lhcb-afb}
      \includegraphics[width=0.45\textwidth]{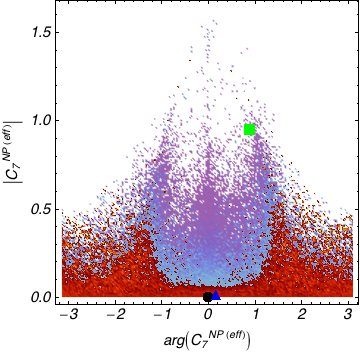}
    }
    \subfloat[~\rate{\Fl}]{
      \label{fig:constraints-lhcb-fl}
      \includegraphics[width=0.45\textwidth]{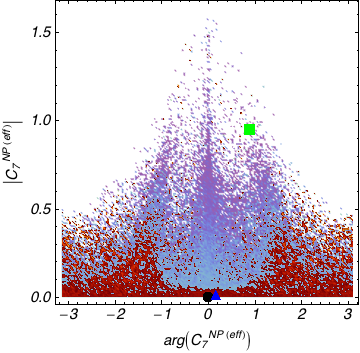}
    }\\
    \subfloat[~\rate{\S{5}} \&  $q^{2}_{0}(\S{5})$]{
      \label{fig:constraints-lhcb-s5}
      \includegraphics[width=0.45\textwidth]{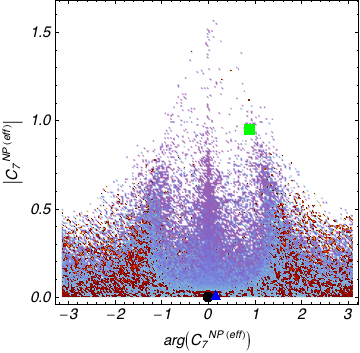}
    }
\caption{\label{fig:constraints_lhcb}[Colour online] The relative impact of different proposed \lhcb measurements after 2\invfb of integrated luminosity, assuming the SM, on the NP component of \Ceff{7}. In each case the colour index has the same meaning as in \fig\ref{fig:constraint-2009-C7}.}
\end{figure}

\fig\ref{fig:constraints_lhcb} shows the relative impact of these measurements on the NP component of \Ceff{7}. In \fig\ref{fig:constraints-lhcb-afb}, SM values of \rate{\Afb} and $q^{2}_{0}(\Afb)$ are imposed with the estimated 2\invfb experimental sensitivities taken from \tab\ref{tab:sense}. \fig\ref{fig:constraints-lhcb-fl} shows the impact of \rate{\Fl}, while \fig\ref{fig:constraints-lhcb-s5} shows the impact of both \rate{\S{5}} and $q^{2}_{0}(\S{5})$ for the same \lhcb integrated luminosity. These should be compared with the currently allowed \Ceff{7} parameter space shown in \fig\ref{fig:constraint-2009-C7}. The small statistical uncertainty found in \seq\ref{sec:sense} for $q^{2}_{0}(\S{5})$ provides a stringent constraint on parameter space. This emphasizes the importance of an early measurement of \S{5}, in addition to \Afb and \Fl.

\begin{figure}
\centering
      \includegraphics[width=0.45\textwidth]{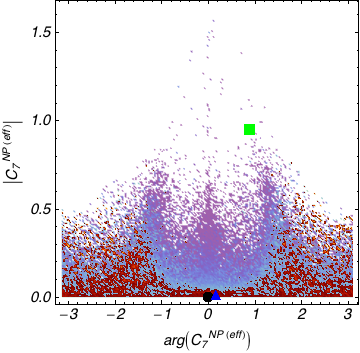}
      \includegraphics[width=0.45\textwidth]{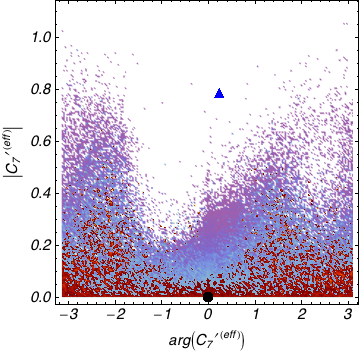}
      \includegraphics[width=0.45\textwidth]{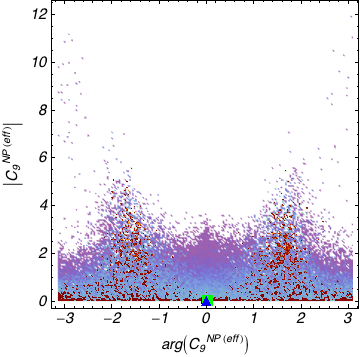}
      \includegraphics[width=0.45\textwidth]{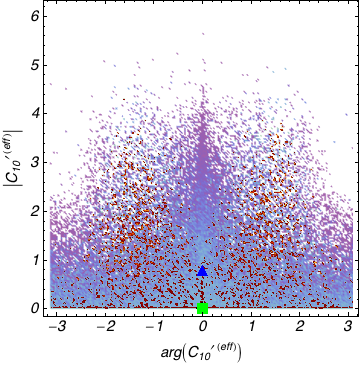}
\caption{\label{fig:constraint-lhcb-all}[Colour online] Allowed parameter space for the Wilson coefficients \Ceff{7}, \Ceffp{7}, \Ceff{9} and \Ceffp{10} after 2\invfb measurements at \lhcb of \rate{\Fl}, \rate{\Afb},  $q^{2}_{0}(\Afb)$, \rate{\S{5}} and  $q^{2}_{0}(\S{5})$, assuming the SM. The colour coding is the same as in \fig\ref{fig:constraint-2009-C7}.}
\end{figure}

\fig\ref{fig:constraint-lhcb-all} shows the combined effect of the measurement of the  proposed observables, again assuming the SM and the estimated sensitivities from \tab\ref{tab:sense} for the NP contribution to the Wilson coefficients \Ceff{7}, \Ceffp{7}, \Ceff{9} and \Ceffp{10}. The amount of parameter space left after these measurements would be significantly reduced, with most NP contributions excluded at the $1\sigma$ level unless there are large NP phases present. This again illustrates the importance of \CP observables as described in \cite{Bobeth:2008ij,Altmannshofer:2008dz}. The FBMSSM and GMSSM models from \seq\ref{sec:model-dependent} could also be excluded at better than 95\% confidence in this case.

\section{Summary}\label{sec:summary}
A new next-to-leading order model of the decay \BdbKsmm, that features QCD factorization corrections and full LCSR form factors, was presented. This includes an expression for the decay amplitude in terms of an updated set of auxiliary functions; these can be compared directly to the previous model, based on \Rf\cite{Ali:1999mm}. The auxiliary functions have been extended to include the effects of primed, scalar, and pseudoscalar operators, which may become important in certain NP scenarios.

The observables \Afb, \Fl, and \S{5} were identified as being promising for a relatively early measurement at the \lhc, as they can be extracted as a function of \qsq by counting signal events in specific angular bins, using \eqs(\ref{eqn:afb_counting})--(\ref{eqn:s5_counting}), and correspond to large features in the angular distribution. We also obtained a simple expression for $q^{2}_{0}(\S{5})$ at leading order, in terms of \Ceff7, \Ceffp7, and \Ceff9, and showed that it has reduced hadronic form factor uncertainties in the large-recoil limit. Considering current experimental constraints leads to restrictions on the possible NP contributions to the Wilson coefficients. The allowed values of the \Afb and \S{5} zero-crossing points, and the gradient of the \Afb and \S{5} distributions at these points, were explored. The relative steepness of the \S{5} distribution, even in the presence of NP, makes $q^{2}_{0}(\S{5})$ an experimentally attractive target, as it will lead to a smaller experimental uncertainty.
 
In order to investigate the impact of measuring the proposed observables on the NP contributions to the Wilson coefficients, and to compare their relative impact, we estimated their sensitivities at \lhcb. We studied the sensitivity to the \qsq integrated values and zero-crossing points of \Afb, \Fl, and \S{5}. The prospect of measuring $S_5$ and its zero-crossing at \lhcb has not been previously explored.

Using a combination of \rate{\Fl}, \rate{\Afb},  $q^{2}_{0}(\Afb)$, \rate{\S{5}}, and  $q^{2}_{0}(\S{5})$, we showed that 2\invfb of \lhcb data could greatly reduce the allowed parameter space. The contribution of \S{5} to this is very significant and can, in part, be attributed to the small statistical uncertainty expected on $q^{2}_{0}(\S{5})$. We have also shown that if the decay is SM-like, the GMSSM and FBMSSM points considered would be ruled out by \lhcb with 2\invfb. We conclude by stressing that making measurements of \S{5} and its zero-crossing would provide an interesting and complementary measurement to others currently planned. \BdbKsmm is a promising channel for constraining models or making a NP discovery. We look forward to the first \lhc results for this decay.
  
\section*{Acknowledgements}
The authors would like to thank: Patricia Ball for many helpful discussions, providing code for the form factors and a careful reading of the manuscript; Ulrik Egede, Mitesh Patel, and Mike Williams for invaluable input into the experimental analysis and comments on the manuscript; Wolfgang Altmannshofer for providing the NP contributions to the Wilson coefficients in the GMSSM and FBMSSM and Adrian Signer for advice concerning quark masses. This work was supported in the UK by the Science and Technology Facilities Council (STFC).
\newpage
\appendix
\section*{Appendix}

\section{Operator Basis}\label{app:WCs}

The effective Hamiltonian for \BdKsmm can be expressed in terms of effective operators and Wilson coefficients as described in \seq{\ref{sec:theory_wilson}}. We provide explicit expressions for a subset of these operators, which play a key role in the decay. Definitions for the remaining operators can be found in \Rf\cite{Bobeth:2008ij}.
\begin{align}
\label{eq:O7}
{\mathcal{O}}_{7} &= \frac{e}{g^2} \bar{m}_b
(\bar{s} \sigma_{\mu \nu} P_R b) F^{\mu \nu} ,&
{\mathcal{O}}_{7}^\prime &= \frac{e}{g^2}  \bar{m}_b
(\bar{s} \sigma_{\mu \nu} P_L b) F^{\mu \nu} ,\\
\label{eq:O8}
{\mathcal{O}}_{8} &= \frac{1}{g} \bar{m}_b
(\bar{s} \sigma_{\mu \nu} T^a P_R b) G^{\mu \nu \, a} ,&
{\mathcal{O}}_{8}^\prime &= \frac{1}{g}  \bar{m}_b
(\bar{s} \sigma_{\mu \nu} T^a P_L b) G^{\mu \nu \, a} ,\\
\label{eq:O9}
{\mathcal{O}}_{9} &= \frac{e^2}{g^2} 
(\bar{s} \gamma_{\mu} P_L b)(\bar{\mu} \gamma^\mu \mu) ,&
{\mathcal{O}}_{9}^\prime &= \frac{e^2}{g^2} 
(\bar{s} \gamma_{\mu} P_R b)(\bar{\mu} \gamma^\mu \mu) ,\\
\label{eq:O10}
{\mathcal{O}}_{10} &=\frac{e^2}{g^2}
(\bar{s}  \gamma_{\mu} P_L b)(  \bar{\mu} \gamma^\mu \gamma_5 \mu) ,&
{\mathcal{O}}_{10}^\prime &=\frac{e^2}{g^2}
(\bar{s}  \gamma_{\mu} P_R b)(  \bar{\mu} \gamma^\mu \gamma_5 \mu) ,\\
\label{eq:OS}
{\mathcal{O}}_{S} &=\frac{e^2}{16\pi^2}
 \bar{m}_b (\bar{s} P_R b)(  \bar{\mu} \mu) ,&
 {\mathcal{O}}_{S}^\prime &=\frac{e^2}{16\pi^2}
 \bar{m}_b (\bar{s} P_L b)(  \bar{\mu} \mu) ,\\
\label{eq:OP}
{\mathcal{O}}_{P} &=\frac{e^2}{16\pi^2}
\bar{m}_b (\bar{s} P_R b)(  \bar{\mu} \gamma_5 \mu) ,&
 {\mathcal{O}}_{P}^\prime &=\frac{e^2}{16\pi^2}
 \bar{m}_b (\bar{s} P_L b)(  \bar{\mu} \gamma_5 \mu),
\end{align}
where $g$ is the strong coupling constant, e is the electron charge, $ \overline{m}_b$ is the $b$ quark mass in the $\overline{\mathrm{MS}}$ scheme, as described in \seq{\ref{sec:theory}}, and $P_{L,R}=(1\mp\gamma_5)/2$.

\section{Angular Coefficients}\label{app:Is}
Here we provide the relations between the angular coefficients, $I^{(s/c)}_i$, defined in \seq\ref{sec:observables} and the auxiliary functions defined in \eq(\ref{eq:AmpAtoI}). We first express the $I^{(s/c)}_i$'s in terms of transversity amplitudes as in \Rf\cite{Altmannshofer:2008dz}. 
\begin{align}
  I_1^s & = \frac{(2+\beta^2)}{4} \left[|\apeL|^2 + |\apaL|^2 + (L\to R) \right] 
            + \frac{4 m_\mu^2}{q^2} \re\left(\apeL^{}\apeR^* + \apaL^{}\apaR^*\right)
\\
  I_1^c & =  |\azeL|^2 +|\azeR|^2  + \frac{4m_\mu^2}{q^2} 
               \left[|A_t|^2 + 2\re(\azeL^{}\azeR^*) \right] + \beta^2 |A_S|^2 ,
\\
  I_2^s & = \frac{ \beta^2}{4}\left[ |\apeL|^2+ |\apaL|^2 + (L\to R)\right],
\\
  I_2^c & = - \beta^2\left[|\azeL|^2 + (L\to R)\right],
\\
  I_3 & = \frac{1}{2}\beta^2\left[ |\apeL|^2 - |\apaL|^2  + (L\to R)\right],
\\
  I_4 & = \frac{1}{\sqrt{2}}\beta^2\left[\re (\azeL^{}\apaL^*) + (L\to R)\right],
\end{align}
\begin{align}
  I_5 & = \sqrt{2}\beta\left[\re(\azeL^{}\apeL^*) - (L\to R) 
- \frac{m_\mu}{\sqrt{q^2}}\, \re(\apaL {A_S^*}+\apaR {A_S^*})
\right],
\\
  I_6^s  & = 2\beta\left[\re (\apaL^{}\apeL^*) - (L\to R) \right],
\\
   I_6^c  &  =
 4 \beta  \frac{m_\mu}{\sqrt{q^2}}\, \re \left[ \azeL {A_S^*} + (L\to R) \right],
\\
  I_7 & = \sqrt{2} \beta \left[\im (\azeL^{}\apaL^*) - (L\to R) 
+ \frac{m_\mu}{\sqrt{q^2}}\, {\im}(\apeL {A_S^*}+\apeR {A_S^*})
\right],
\\
  I_8 & = \frac{1}{\sqrt{2}}\beta^2\left[\im(\azeL^{}\apeL^*) + (L\to R)\right],
\\
  I_9 & = \beta^2\left[\im (\apaL^{*}\apeL) + (L\to R)\right].
\end{align}
These transversity amplitudes are projections of the decay amplitude onto various combinations of helicity states of the \kstar and the virtual gauge boson. The projections can be achieved by contracting $\mathcal{T}_\mu^{1/2}$ with the virtual gauge boson polarization vector.  We use four basis vectors for the virtual gauge boson polarization vector corresponding to transverse ($\pm$), longtitudinal (0) and time-like (t) states, and three basis vectors for the virtual gauge boson polarization vector corresponding to transverse ($\pm$) and longtitudinal (0) states. One first extracts the helicity amplitudes $H_+$, $H_-$ and $H_0$ using the basis polarization vectors +,-,0 respectively for both the \kstar and the virtual gauge boson. $H_t$ is found by taking the longtitudinal polarization vector for the \kstar and the time-like polarization vector for the virtual gauge boson. Using the relations 
\begin{equation}
 A_{\perp/\parallel}=\frac{H_{+}\mp H_{-}}{\sqrt{2}}
\end{equation}
and $A_0=H_0$, $A_t=H_t$, one then obtains expressions for the transversity amplitudes in terms of $A(q^2)$ to $S(q^2)$,
\begin{align}
A^{i}_\perp(q^2)&=\sqrt{2\, \lambda}\,N\,m_B\,c_{i}(q^2)\\
A^{i}_\parallel(q^2)&=-\sqrt{2}\, N\,m_B\,a_{i}(q^2)\\
A^{i}_0(q^2)&=\frac{N\,m_B}{\hat{m}_{K^*} \,\sqrt{\hat{q}^2}}\left(-\frac{1-\hat{m}_{K^*}^2-\hat{q}^2}{2}\,a_{i}(q^2)+\lambda \,b_{i}(q^2) \right)\\
A_t(q^2)&=\frac{N\,m_B\, \sqrt{\lambda}}{\hat{m}_{K^*}\,\sqrt{\hat{q}^2}} \left(F(q^2)-(1-\hat{m}_{K^*}) G(q^2)-\hat{q}^2 H(q^2)\right),
\end{align}
where $i=L/R$. We use the standard normalization and definitions following \Rf\cite{Kruger:2005ep},  
\begin{align}
 \beta=&\sqrt{1-\frac{4 m_\mu^2}{q^2}}\\ \lambda=&1+\hat{m}_{K^*}^4+\hat{q}^4-2\left(\hat{q}^2+ \hat{m}_{K^*}^2\left(1+\hat{q}^2\right)\right)\\
N=&\left(\frac{G_F^2\, \alpha^2}{3\cdot 2^{10} \pi^5\, m_B} |V_{ts} V^*_{tb}|^2 q^2\, \lambda^{1/2}\,\beta\right)^{\frac{1}{2}},
\end{align}
where $\alpha$ is the electromagnetic coupling constant and $G_F$ is the Fermi constant. In the above definitions of the transversity amplitudes, the functions $a_{L/R}(q^2)$, $b_{L/R}(q^2)$, $c_{L/R}(q^2)$, are analogous to those defined in \Rf\cite{Kim:2000dq},
\begin{align}
a_{L/R}(q^2)&=B(q^2)\mp F(q^2),\\
b_{L/R}(q^2)&=\frac{1}{2} \left(C(q^2)\mp G(q^2)\right),\\
c_{L/R}(q^2)&=\frac{1}{2} \left(A(q^2)\mp E(q^2)\right).
\end{align}
Using the above it is possible to compare the predictions of \eqs(\ref{eq:AmpAtoI}) to the standard results in the literature, and we agree with \Rf\cite{Altmannshofer:2008dz}. 
\bibliographystyle{JHEP}
\bibliography{references}
\end{document}